\documentclass[useAMS,usenatbib]{mn2e}
\usepackage{graphicx}
\usepackage{times}
\usepackage{url}

\newcommand{\rr}{53}
\newcommand{\ea}{142}
\newcommand{\eb}{23}
\newcommand{\ew}{218}
\newcommand{\ceph}{26}
\newcommand{\puls}{153}
\newcommand{\lpv}{222}
\newcommand{\spot}{13}
\newcommand{\skycov}{$\sim150$ deg$^2$}
\newcommand{\newd}{659}
\newcommand{\total}{850}  
\newcommand{\fields}{25}
\newcommand{\totlcs}{$\sim87000$}

\newcommand{\dg}{\mbox{$^\circ$}}

\newcommand{\as}{\mbox{$^{\prime\prime}$}}
\newcommand{\dsc}{$\delta$ Scuti}

\newcommand{\aj}{AJ}

\newcommand{\aaa}{A\&A}

\newcommand{\aas}{A\&AS}

\title[UNSW variable star catalogue]{The University of New South Wales Extrasolar Planet Search: a catalogue of variable stars from fields observed 2004--2007}
\author[J. L. Christiansen et al.]{
	J. L. Christiansen$^{1}$\thanks{E-mail:jessiec@phys.unsw.edu.au (JLC)}, 
	A. Derekas$^{2}$,
	L. L. Kiss$^2$,
	M. C. B. Ashley$^{1}$,
	S. J. Curran$^{1}$,
	\newauthor
	D. W. Hamacher$^{1}$,
	M. G. Hidas$^{3,4}$,
	M. R. Thompson$^{5}$,
	J. K. Webb$^{1}$,
	and T. B. Young$^{1}$\\
$^{1}$School of Physics, University of New South Wales, Sydney 2052, Australia\\
$^{2}$School of Physics, University of Sydney, Sydney 2006, Australia\\
$^{3}$Las Cumbres Observatory Global Telescope, Goleta, CA 93117, USA\\
$^{4}$Department of Physics, University of California, Santa Barbara, CA 93106, USA\\
$^{5}$High Performance Computing Support Unit, University of New South Wales, Sydney 2052, Australia}

\begin{document}

\date{\today}

\pagerange{\pageref{firstpage}--\pageref{lastpage}} \pubyear{2007}

\maketitle

\label{firstpage}

\begin{abstract}

We present a new catalogue of variable stars compiled from data taken for the University of New
South Wales Extrasolar Planet Search. From 2004 October to 2007 May, \fields~target fields were
each observed for 1--4 months, resulting in \totlcs~high precision light
curves with 1600--4400 data points. We have extracted a total of \total~variable light curves,
\newd~of which do not have a counterpart in either the General Catalog of Variable Stars, the New Suspected Variables catalogue or the All Sky Automated Survey southern variable star catalogue. The catalogue is detailed here, and includes \ea~Algol-type eclipsing binaries, \eb~$\beta$ Lyrae-type eclipsing binaries, \ew~contact eclipsing binaries, \rr~RR Lyrae stars,
\ceph~Cepheid stars, \spot~rotationally variable active stars, \puls~uncategorised pulsating stars with periods
$<$10~d, including \dsc~stars, and  \lpv~long period variables with variability on timescales of $>$10~d. As a general application of
variable stars discovered by extrasolar planet transit search projects, we discuss several 
astrophysical problems which could benefit from carefully selected samples of bright variables.
These include: {\it (i)} the quest for contact binaries with the smallest mass ratio, which could be used
to test theories of binary mergers; {\it (ii)} detached eclipsing binaries with pre-main-sequence
components, which are important test objects for calibrating stellar evolutionary models; and
{\it (iii)} RR~Lyrae-type pulsating stars exhibiting the Blazhko-effect, which is one of the last
great mysteries of pulsating star research. 

\end{abstract}


\section{Introduction}
\label{sec:int}

The University of New South Wales (UNSW) is conducting a wide-field survey for transiting extrasolar planets, and is one of an increasing number teams around the world using this method. The nature of wide-field surveys has resulted in an enormous number of high-precision light curves being produced, numbering in the millions for some teams (e.g. \cite{Cameron07}). 

In order to maximise the output efficiency from wide-field surveys, it is important to make the data available for use in other studies once planet candidates have been identified. The most extensive results produced by these projects to date have been long lists of newly discovered variable stars, inevitably with very limited information apart from the period and amplitude in a single band \citep{Hartman04,Pepper06}. Therefore, one can imagine the main use of these variable star catalogues is to define starting samples for astrophysically interesting follow-up studies that benefit from large samples of carefully selected stars. A recent example is the list of variable stars coincident with x-ray sources presented by \cite{Norton07}.

The UNSW Extrasolar Planet Search is performed with the largest clear aperture telescope of the wide-field transit surveys. With a diameter of 0.5-m, this project occupies the niche between the typical wide-field transit surveys observing brighter targets with 0.1--0.2-m diameter telescopes, and the deeper surveys with narrower fields of view using $>$1-m diameter telescopes. The larger collecting area in this project has been exploited to increase the acquisition rate for the observations, as compared to observing deeper targets, since brighter targets have a higher potential for interesting follow-up studies. The large data set of light curves we have obtained is therefore the ideal starting point from which to compile a bright variable star catalogue of particularly well-sampled light curves with high precision photometry and moderately long observing baselines (1--4 months).

The paper is organised as follows. Section \ref{sec:obs} describes the observations and reduction pipeline. Section \ref{sec:sel} describes the methods by which the variable light curves were selected. The final catalogue is presented in Section \ref{sec:res}, while three possible applications of the sample are discussed in Section \ref{sec:dis}. The data are publicly available at the University of New South Wales Virtual Observatory (VO) facility, which is described in Section\ \ref{sec:online}. We close the main body of the paper with a  short summary of the project in Section\ \ref{sec:summary}. Cross-references to the General Catalogue of Variable Stars (GCVS) and the All Sky Automated Survey (ASAS) database are given in the Appendix.

\section{Observations and Reduction}
\label{sec:obs}

\subsection{Photometry}
\label{sec:phot}

The data were obtained using the dedicated 0.5-m Automated Patrol Telescope at Siding Spring Observatory, Australia. Observing is performed remotely on every clear night when the Moon is not full and is almost entirely automated; the observer initiates the observing script and monitors the weather conditions. The CCD camera used for these observations consists of an EEV CCD05-20 chip, with $770\times1150$ pixels. The pixel size of 22.5~$\mu$m produces a relatively low spatial resolution of 9.4~arcsec pixel$^{-1}$, and the field of view of each image is $2\times3$~deg$^2$. The observations were taken through a Johnson $I$ filter, a decision designed to maximise the contribution to the photometry of later spectral type dwarf stars, around which it is easier to detect transiting planets. A new CCD camera covering $7\times7$~deg$^2$ with a higher spatial resolution of 4.19~arcsec pixel$^{-1}$ has been constructed for this project and will be installed on the telescope in 2008.

Observations were obtained for 32 months from 2004 October to 2007 May on 25 target fields, listed in Table \ref{tab:fields}, resulting in a total sky coverage of \skycov. The equatorial coordinates of
the centre of each field is given, as are the Galactic coordinates. The strategy
employed for field selection was again motivated by transit detection. The most southerly fields were chosen in order to reduce the airmass variations
over the course of the night, with a maximum allowable declination of 70\dg~due to building
constraints. At the same time, the Galactic latitude was constrained to $b>10$\dg~to alleviate crowding effects in the field, which led to several more northern fields being selected. This latter constraint was relaxed for the final field in order to observe a more crowded stellar field.

Most of the fields were observed in pairs in order to increase the number of target stars, with observations alternating between the two fields over the course of the night. For the majority of the fields the rate of acquisition is 15 images per hour, however for the first pair of fields it is half as often as this. For the final three fields we implemented an automated script that adjusted the exposure times according to the sky brightness levels, and the rate of acquisition ranges from 10--40 images per hour. These fields were also observed singly instead of pairwise, which accounts for the higher rates achieved. Each field was observed for a minimum of 20 nights for at least $\sim4-5$ hours per night, resulting in $1600-4400$ observed data points for each star. Each field contained $\sim1200-8000$ stars with $8.0 \ge I \ge 14.0$, depending on the Galactic coordinates. The numbers of stars observed down to 14${^{\rm th}}$ magnitude in each field are included in Table \ref{tab:fields}, with \totlcs~light curves being generated. 

\begin{table}
 \centering
  \caption{Details of the target fields observed with the Automated Patrol Telescope. The columns for each field are as follows: the field name, the coordinates and Galactic latitude of the centre of the field, the total observations obtained, the number of stars in each field brighter than $I = 14^{\rm th}$ magnitude, and the number of images obtained per hour.}
 \label{tab:fields}
  \begin{tabular}{@{}lllrlll@{}}
  \hline
  Field       & RA        & Dec           & $b$\dg & Obs    &  Stars & Image rate \\
              & (J2000.0)   & (J2000.0)       &        &        &        & $({\rm hr}^{-1})$\\
  \hline
  L1	      &  04~56~24 & $-30$~00~00   & $-36.8$  & 1791    & 2195  & 7.5	 \\
  L2	      &  04~45~00 & $-26$~18~00   & $-38.3$  & 1692    & 1943  & 7.5	 \\
  N1	      &  09~05~00 & $-14$~30~00   & $21.1$   & 1824    & 3271  & 15	 \\
  N2	      &  09~25~00 & $-13$~30~00   & $25.5$   & 1619    & 2495  & 15	 \\
  O3	      & 12~00~00 & $-36$~00~00  & $25.7$   & 2441	& 3066  & 15	 \\
  O4	      & 12~00~00 & $-38$~10~00  & $23.6$   & 2420	& 2008  & 15	 \\
  Q1	      & 17~06~00 & $-60$~00~00  & $-11.4$  & 2083	& 8096  & 15	 \\
  Q2	      & 17~09~00 & $-57$~55~00  & $-10.5$  & 1854	& 7559  & 15	 \\
  R1	      &  00~00~00 & $-59$~00~00  & $-56.9$  & 3708    & 1401  & 15	 \\
  R2	      &  00~00~00 & $-57$~00~00  & $-58.8$  & 3446    & 1496  & 15	 \\
  S5	      &  04~03~00 &  $-02$~55~00  & $-38.3$  & 1980    & 1327  & 15	 \\
  S6	      &  04~06~00 &  $-04$~55~00  & $-38.7$  & 1944    & 1284  & 15	 \\
   Jan06\_1   &  09~20~00 & $-24$~30~00  & $17.5$   & 1713    & 3261  & 15	 \\
   Jan06\_2   &  09~15~00 & $-22$~30~00  & $17.9$   & 1773    & 3520  & 15	 \\
   Feb06\_1   & 12~55~00 & $-45$~15~00  & $17.6$   & 2732    & 4734  & 15	 \\
   Feb06\_2   & 13~15~00 & $-45$~10~00  & $17.5$   & 2631    & 4975  & 15	 \\
   Apr06\_1   & 14~48~00 & $-39$~00~00  &  $18.5$  & 1840    & 5176  & 15	 \\
   Apr06\_2   & 14~48~00 & $-41$~00~00  &  $16.7$  & 1840    & 3996  & 15	 \\
   May06\_1   & 18~27~00 & $-65$~00~00  & $-21.9$  & 2579	& 4421  & 15	   \\
   May06\_2   & 18~27~00 & $-67$~00~00  & $-22.5$  & 2731	& 4269  & 15	   \\
   Jul06\_1   & 21~09~00 & $-66$~30~00  & $-38.2$  & 2065    & 2246  & 15	 \\
   Jul06\_2   & 21~09~00 & $-68$~30~00  & $-37.5$  & 2034    & 2303  & 15	 \\
   Sep06      & 23~42~00 & $-69$~24~00  & $-46.5$  & 3497    & 1630  & 10--40   \\
   Dec06      &  08~03~00 & $-67$~24~00  & $-18.4$  & 3007    & 3387  & 10--40   \\
   Mar07      & 14~15~00 & $-69$~00~00  & $-7.3$   & 4450	& 6714  & 10--40   \\
\hline
\end{tabular}
\end{table}

\subsection{Reduction pipeline}
\label{sec:red}

In order to achieve the extremely precise photometry required for transit detection, we
have developed a simple, robust, automated aperture photometry reduction pipeline. A
detailed description can be found in \cite{Hidas05}; a summary is included here for
completeness, and there have been no modifications.

Using the tools developed by \cite{Irwin01}, each image is processed in the standard manner, including bias subtraction, flat-field correction
and catalogue generation. For each field, a master image is generated by combining $\sim10$ consecutive, low airmass images with small
image-to-image shifts, and a master coordinate list is produced. Each image frame is then transformed into the master reference frame with a
positional accuracy of better than 0.01 pixels. Aperture photometry is performed on the transformed images, with a fixed aperture radius
of 3~pixels, which equals 30\as on the sky. At the typical Galactic latitude of our fields, this results in multiple stars falling within the
same photometry aperture more than 80\% of the time; these stars can only be resolved in higher spatial resolution images. As a result, the
magnitudes listed in this catalogue should be taken as upper limits on the true magnitude, and the variability amplitudes as lower limits.
Magnitude variations from image to image are calibrated by using a subset of the brightest stars. The magnitude residuals $\Delta m$ for each
star are fit iteratively with a position-dependent function of the form

\begin{equation}
  \Delta m = a + bx + cy + dxy + e^2
\label{eq:cal}
\end{equation}

\noindent where $x$ and $y$ are the pixel coordinates of the star on the CCD, and $a-e$ are a set of constants for each image. With each iteration, the stars with the highest r.m.s. residuals are removed.

\begin{figure}
\begin{center}
\includegraphics[width=8.5cm]{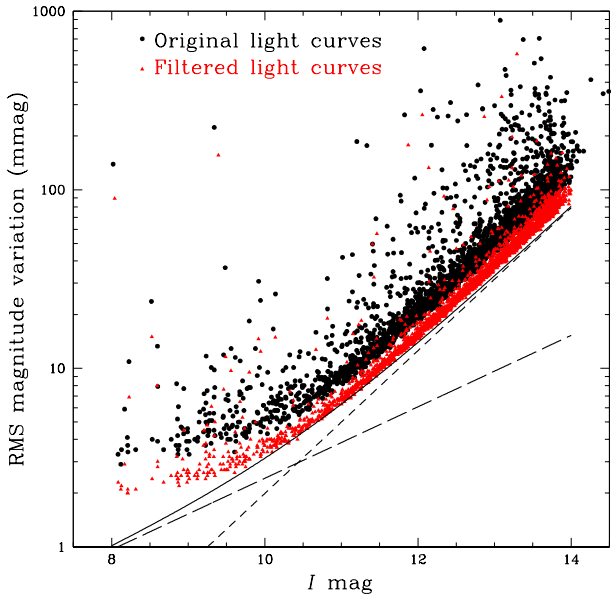}
\caption{\small A typical example of the precision of the photometry we have obtained with
our reduction pipeline and additional trend-filtering to remove systematic noise. The
target field is Jan06\_1 and light curves are comprised of 1713 data points. The solid circles are the original light curves produced by the
pipeline; the hollow triangles show the improvement in the precision with the
trend-filtering. The solid line shows the theoretical Poisson noise limits from the sky flux (short dashed line) and stellar flux (long dashed line).}
\label{fig:precision}
\end{center}
\end{figure}

After the light curves are generated by the reduction pipeline, they are processed to remove the significant systematic signals using an implementation of the trend-filtering algorithm described by
\cite{Kovacs05}. A random subset of several hundred stars are chosen as
template light curves. For each of the remaining light curves, the closest matching synthetic light curve that can be 
reconstructed from a linear combination of the template light curves is subtracted. Signals that are
common to the template and original light curve will be removed, and signals that are
unique to the original light curve remain. Figure \ref{fig:precision} shows an example of
the precision achievable in a typical field of data, before and after being processed with the trend-filtering algorithm. For the stars brighter than $I\sim11.5$ the r.m.s. precision of the filtered light curves is less than 10 mmag.

\section{Selection of variable candidates}
\label{sec:sel}

Three methods were used to extract the variable light curves from the full data set: {\it (i)} visual inspection of the filtered light curves down to 13$^{\rm th}$ magnitude, {\it (ii)} implementation of a box-search algorithm on all filtered light curves (down to 14$^{\rm th}$ magnitude), and {\it (iii)} implementation of the Stetson Variability Index on the entire set of filtered and unfiltered light curves. The first two methods formed part of the search for transiting extrasolar planets, the main science driver of this project \citep{Hidas05}. The third was implemented to improve the completeness of the catalogue.

\subsection{Visual inspection}
\label{sec:vis}

As a first pass, all light curves down to 13$^{\rm th}$ magnitude are visually inspected. This has the advantage of being reasonably immune to the effects of systematic variability in the light curves, since the brain can be quickly trained to filter out similar signals appearing in multiple light curves. However, it becomes less useful for light curves fainter than 13$^{\rm th}$ magnitude due to the increased shot noise; it can also fail for brighter stars where the variable signal is of low significance, and will not be evident until the light curve is phased with the correct period. Importantly, this method is not successful for the detection of variable light curves with periods greater than $\sim$5~d. In these cases, the light curve for each night appears essentially flat, especially to someone specifically searching for transit events on the order of a few hours. Also, visual inspection has the potential to miss light curves that exhibit only single or partial events on a single pass through.

\subsection{Transit detection algorithm}
\label{sec:trans}

The light curves are subsequently processed with a transit detection algorithm \citep{Aigrain04} which searches for box-shaped transit events within specified transit duration and period windows. For each light curve, the algorithm determines the combination of epoch, transit duration and period within a specified range that returns the highest signal-to-noise (S/N) ratio. Subsequently, light curves with a S/N greater than some cut-off (typically $>$8.0, although this value varied with the degree to which each field was affected by systematics) when folded at the these parameters are visually inspected in both raw and folded formats. We have chosen the transit duration and period windows for maximum efficiency in the extrasolar planet transit search, using windows of 0.04--0.25~d and 1.0--5.0~d respectively. The algorithm will detect the variable light curves with parameters that fall within these windows: both the shallow transit events that are flagged as potential transiting planet candidates and the deeper events produced by detached eclipsing binary systems. Additionally, it returns many variable light curves which can be approximated to some extent by a box-shaped model lying within the required period range, including: grazing eclipsing binaries exhibiting V-shaped transits; continuously varying light curves that give a significant result when folded to an optimum period; and variable light curves with periods outside the specified window but where an integer number of periods is located within the window. It also has the additional advantage of detecting those light curves with only single or partial events and providing a potential period. However, it is not useful for detecting variable light curves with periods much shorter ($<$0.5~d) or much longer ($>$10~d) than the specified period window. 

\subsection{Stetson Variability Index}
\label{sec:svi}

Neither of the preceding methods will rigorously detect the longest period variables in our light curves. In order to increase the completeness of this variable star catalogue it was essential to correct this bias. An additional method of detecting variability in light curves is the Stetson Variability Index \citep{Stetson96}, a measure of the correlated signal in a light curve.

\begin{figure}
\begin{center}
\includegraphics[width=8.5cm]{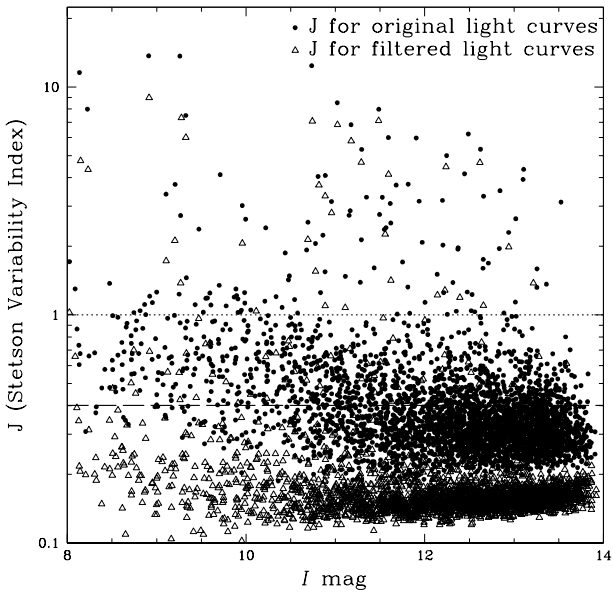}
\caption{\small The Stetson Variability Index $J$ as a function of magnitude for the Dec06
field. The solid circles are the original light curves, and the hollow triangles are the
filtered light curves. During the variable selection process unfiltered light curves with
$J>1.0$ (the dotted line) and filtered light curves with $J>0.4$ (the dashed line) were
flagged.}
\label{fig:js}
\end{center}
\end{figure}

Using the notation of \cite{Stetson96}, the index $J$ is given by

\begin{equation}
  J=\frac{\sum_{k=1}^n w_k{\rm sgn}(P_k)\sqrt{|P_k|}}{\sum_{k=1}^n w_k}
\label{eq:js}
\end{equation}

\noindent where $n$ pairs of observations have been defined. For the $k$th pair, with a weighting of
$w_k$, the magnitude residuals are $\delta_{i(k)}$ and $\delta_{j(k)}$, where $i$ and $j$ are the observations forming the pair. We can therefore define the product of the magnitude residuals as $P_k = \delta_{i(k)}\delta_{j(k)}$, or $P_k = \delta_{i(k)}^2-1$ for single observations (where $i = j$). The term ${\rm sgn}(P_k)$ is the sign (positive or negative) of $P_k$. We have calculated the
magnitude residuals in the same manner as \cite{Stetson96}, scaling by the individual
observational errors and correcting for the statistical bias to the mean, giving

\begin{equation}
  \delta = \sqrt{\frac{n}{n-1}}\left(\frac{\upsilon - \bar \upsilon}{\sigma_{\upsilon}}\right)
\label{eq:delta}
\end{equation}

\noindent where $\upsilon$ is the measured magnitude of the observation, $\bar \upsilon$ is the mean magnitude over all observations, and $\sigma_{\upsilon}$ is the individual error on the observed magnitude. To form the pairs of observations we chose a timescale of 10 minutes; all observations
that lie within 10~minutes of each other are paired. All pairs with $i \neq j$ are assigned
a weight of 1.0, and those with $i = j$ a weight of 0.1. We found the best results in terms
of detecting longer period variables ($>$ 1 d) when the data were binned on a similar
timescale of $\sim$10 minutes, although this was at the cost of lowering the detection of
the very shortest period variables. The solid squares in Figure \ref{fig:js} show a typical
distribution of this variability index for a single field, in this case the Dec06 field as
shown in Figure \ref{fig:precision}, prior to the trend-filtering stage.

One problem we encountered was the tendency for our implementation of the trend-filtering
algorithm to suppress or entirely remove the night-to-night magnitude jumps present in the
long-period variable light curves, resulting in a smaller than expected variability index.
This was solved by running the variability index on both the filtered light curves to
detect the shorter period variables and the unfiltered light curves to detect the longer
period variables. The caveat to this is that long period trends in the systematic signals
in the data, for instance signals correlated with moon phase, are not removed from the long
period light curves. In an effort to overcome this, we have removed those long period
light curves where multiple light curves in the same field demonstrate the same morphology
and are described well by the same period and epoch. In the future we plan to resolve this
problem by replacing the trend-filtering algorithm with an implementation of SYSREM
\citep{Tamuz05}. Preliminary tests indicate this will not affect the longer period
variable light curves in the same manner. The hollow triangles in Figure \ref{fig:js} show
the distribution of the variability index after trend-filtering. For the unfiltered light
curves, we set a cut-off of $J = 1.0$, and for the filtered light curves we set $J = 0.4$.
We found these limits recovered 90--97\% of the variables previously identified by visual
inspection and the box-search algorithm, as well as over 150 long-period variables that had previously been undetected. The variables that were not recovered were generally the shallower longer period eclipsing binary light curves where
the occasional small excursion from the mean magnitude was not sufficient to increase the variability
index above the cut-off. Also missing were the shortest period variables with periods less than 1
hr, where the timescales for pairing and binning of 10 minutes were long enough to reduce
the effectiveness of the variability index as a true measure of variability.

\section{The light curve catalogue of variable stars}
\label{sec:res}

Using these methods, we find a total of \total~variable light curves in our data set. These
have been analysed in a similar iterative fashion to Derekas et al. (2007) as follows. 
Initial periods have been determined with either the transit detection algorithm (for eclipsing light curves)
or $\chi^2$ fitting of sine waves using discrete Fourier transforms (for continuously varying light curves). Each of the resulting phased
curves was then visually inspected to assign a type of variability, and also to confirm that the automatically determined period was not half
or an integer multiple of the real period, a common occurrence for light curves of eclipsing binaries. A visual inspection of every phase
diagram was usually sufficient to show whether the determined period was an alias or was slightly inaccurate. In the case of an alias, we
multiplied the initial period by different constants (in most cases by 2) until the shape of the curve was consistent with that of an
eclipsing binary. For the long period variables, the observing baseline was typically insufficient to determine if the variability was
periodic; in these cases the period and epoch are not supplied in the catalogue.

We next used the string-length method (Lafler \& Kinman 1965; Clarke 2002) to improve the
period determination (see also Derekas et al. 2007 for further details). We applied the method
for 500~periods within $\pm1$\% of the best initial period guess. The typical period
improvement resulted in a change in the 3rd-4th decimal place, consistent with the
limited frequency resolution of the data (which scales with 1/T$_{\rm obs}$, where T$_{\rm
obs}$ is the time-span of the observations). 

During the individual inspection of the phase diagrams, we made a visual classification of all
\total~variables. Based on the light curve shapes alone, phased with the final adopted periods,
we placed each star into one of the following categories: Algol-type (EA), $\beta$ Lyrae-type
(EB), W Ursae Majoris-type (EW), RR~Lyrae stars (RRL), Cepheids (DCEP), long period variables
with periods $>10$ d (LPV), and pulsating variables with periods $<10$ d (including $\delta$
Scuti and other multiply periodic variables, referred to as PUL). We follow the convention of
using a colon to indicate a loose classification (for example, EB:). In several cases we used
the ``spotted variable'' type, which refers to singly periodic variables with periods of several
days, light curve amplitudes of a few hundredths of a magnitude and light curve shapes
characteristic of known rotationally variable active stars. These can be binaries or single
stars, and have multi-periodic light variations on time scales of years and decades (see for
example Ol\'ah et al. 2000). We note that for sinusoidal light curve shapes, it is
difficult to differentiate between the EW, PUL and spotted variable classifications by eye.
Where there is an ambiguity between several classifications they are listed as, for example,
EW/PUL. If there are two types of variability present they are listed as, for example, EA+PUL.
If there is additional information it is given as, for example, RRL-Blazhko.

\begin{figure}
\begin{center}
\includegraphics[width=8.5cm]{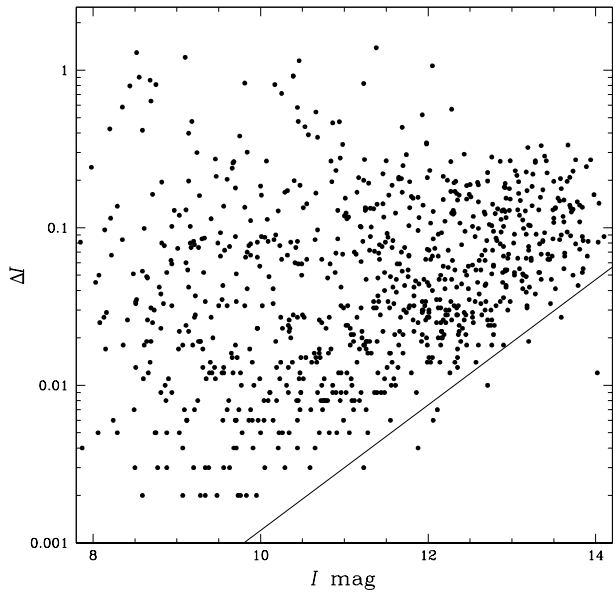}
\caption{\small The variability amplitude detection limits for this catalogue. The mean $I$-band
magnitude and the variability amplitude are plotted for the entire data set of variable light
curves. }
\label{fig:limits}
\end{center}
\end{figure}

The detection limits of the catalogue are shown in Figure \ref{fig:limits}, with the mean
$I$-band magnitude and variability amplitude plotted for all \total~variable light curves. For
the detached eclipsing binaries the amplitude was the best-fitting transit depth as recorded
by the box-search algorithm; for the continuously varying light curves we have used the
amplitude from the sinusoid-fitting. For the multi-periodic light curves this will represent an approximate amplitude of the dominant
frequency. As discussed in Section \ref{sec:red}, due to dilution of the signal in crowded
photometry apertures, the variability amplitudes presented here are lower limits on the true
amplitudes. From Figure \ref{fig:limits} it is apparent that around $I\approx12$ mag we lose sensitivity
to the lowest amplitude variables (such as the multi-periodic \dsc~stars or pulsating
red giants), while for $I>13$ mag only the highest amplitude pulsators (RR~Lyrae stars) and
eclipsing binaries remain. 

The last step in the variable star analysis was cross-correlation with existing databases to supplement the catalogue with as much additional
information as possible. Namely, we queried the most recent update of the General Catalogue of Variable Stars (GCVS, Samus et al. 2007),
including the New Suspected Variables catalogue, to identify already known variable stars. In addition, we checked the ASAS-3 database of
southern variables (Pojmanski 2002). This revealed that 191 out of \total~variables are positionally coincident with previously published
variable stars, leaving the total number of our new discoveries at \newd.  This corresponds to 78\%, which is a lower fraction than, for
instance, the 90\% new discoveries found by Hartman et al. (2004) in the HATNet observations of the Kepler field. However, it is still
surprisingly large, given the fact that the ASAS-3 project had previously observed each of our fields, whereas the Kepler field had not been
targeted with variability surveys prior to the Hartman et al. study. We also performed a cross-correlation with the 2MASS Point Source Catalog
(Skrutskie et al. 2006) to provide $JHK$ magnitudes. Where multiple 2MASS sources are present within the APT photometry aperture, the source
closest to the centre of the photometry aperture that is brighter than $J\sim15$~mag is selected. Finally the catalogue was cross-correlated
with the ROSAT X-Ray Source Catalog \citep{Voges99,Voges00} to determine which sources, if any, might be active stars with hot coronae. 

\begin{figure}
\begin{center}
\includegraphics[width=8.5cm]{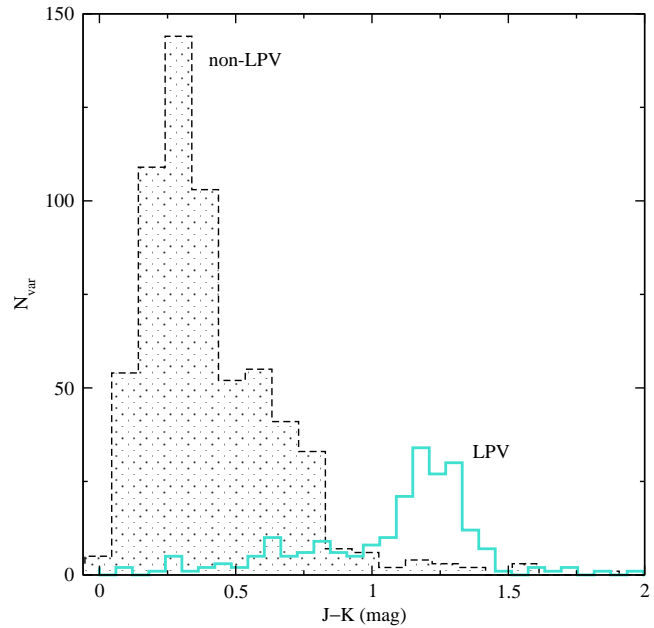}
\caption{\small A comparison of the colour histograms for LPV and non-LPV variables.}
\label{fig:jkhist}
\end{center}
\end{figure}

A histogram of the $J-K$ colour indices (Figure \ref{fig:jkhist}) for the LPV and non-LPV
variables (the latter including all eclipsing binaries and classical pulsators) demonstrates the
expected dichotomy, with LPVs mostly having $J-K>0.6$ mag, i.e. being red giant stars. A few
LPVs have bluer colours, which might indicate early-type stars with longer periods unrelated
to red giant pulsations (e.g. ellipsoidal variability in binaries, rotational modulation due to
starspots), or mismatches with the 2MASS catalogue. Conversely, most of the non-LPVs have $J-K<0.7$ mag, corresponding to spectral types
A--K. The few redder non-LPVs are all located at lower galactic latitudes,
suggesting strong interstellar reddening in their cases.

\begin{figure}
\begin{center}
\includegraphics[width=8.5cm]{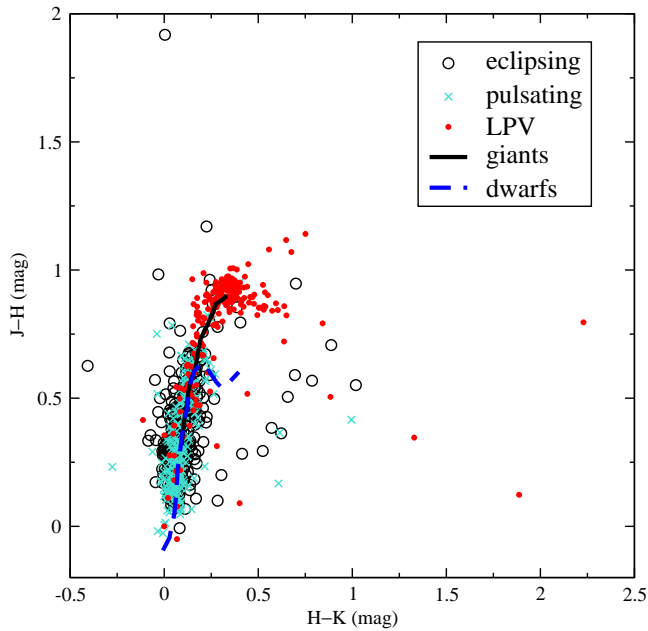}
\caption{\small The $J-H$ vs. $H-K$ colour-colour diagram with the three broad categories
and the stellar loci taken from Bessell \& Brett (1988).}
\label{fig:2cols}
\end{center}
\end{figure}

We also tested the consistency between the assigned variability types and their expected 
stellar types via the $J-H$ vs. $H-K$ colour-colour diagram. Using the intrinsic stellar loci 
determined for dwarfs and giants by Bessell \& Brett (1988) and transformed into the 2MASS
system (Carpenter 2001), we plot the locations of stars in three broad categories (eclipsing,
pulsating, LPV) in Figure \ref{fig:2cols}. Here we find a good agreement: almost all LPVs follow
the intrinsic location of red giant stars, even showing hints of the separate  carbon-rich LPV
sequence for $J-K>1.0$ mag and $H-K>0.4$ mag. There are several outliers towards both bluer
and redder $H-K$ colours, almost exclusively eclipsing binaries, where we may suspect high
reddening, mismatches with the 2MASS catalogue, composite colours of stars blended in the 2MASS catalogue, which has a spatial
resolution of 1~arcsec~pixel$^{-1}$, or large photometric errors in the 2MASS magnitudes.

\begin{figure*}
\begin{center}
\includegraphics[width=14cm]{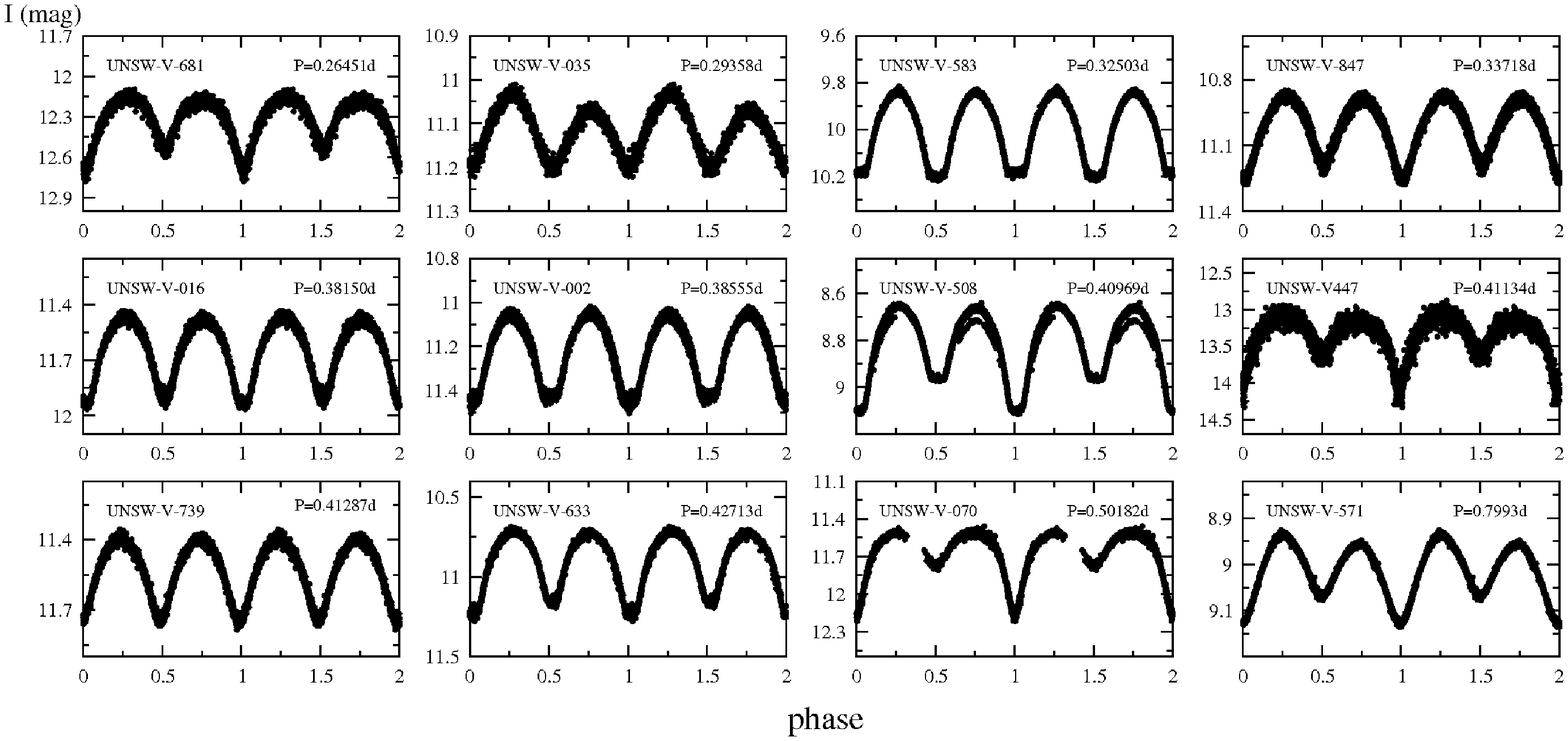}
\includegraphics[width=14cm]{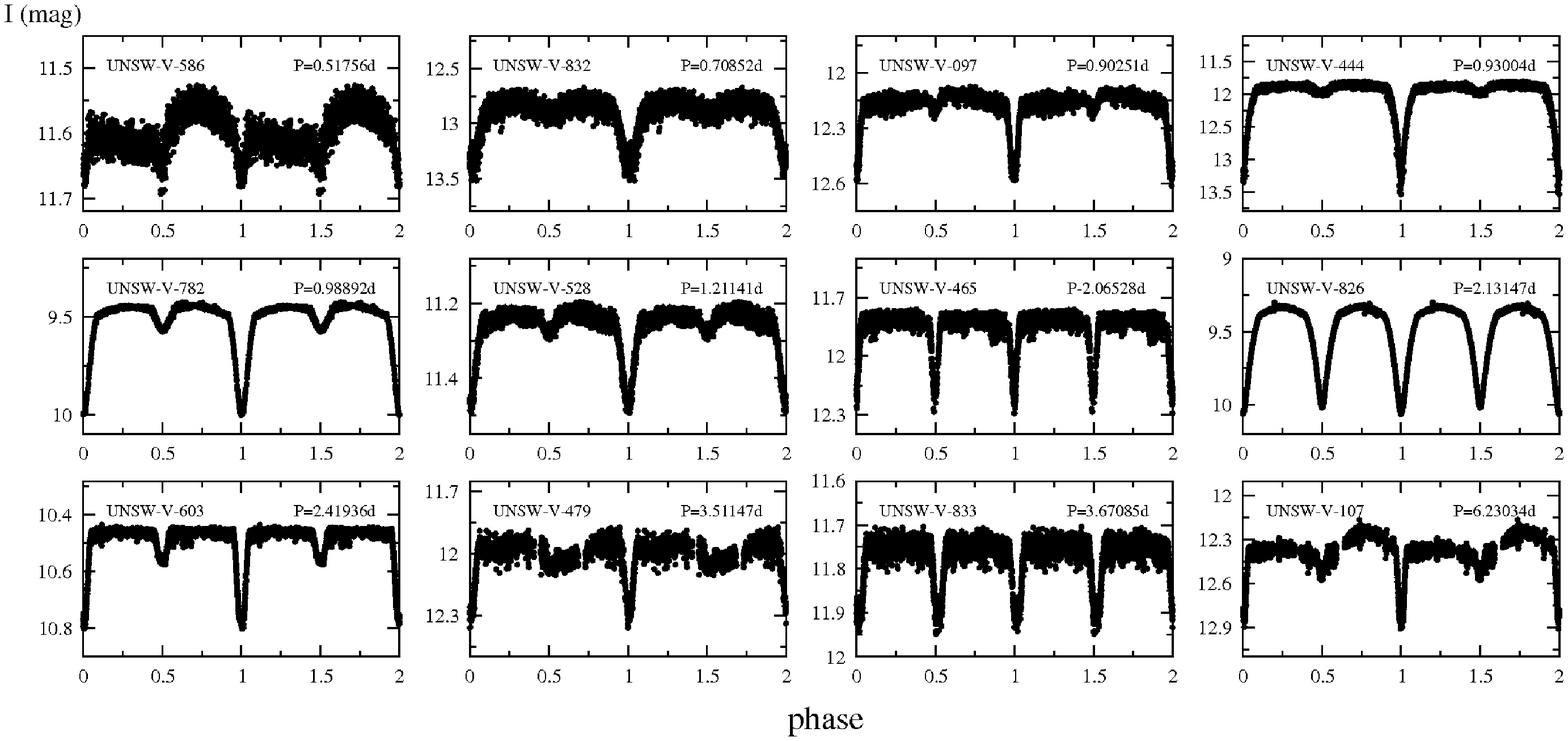}
\includegraphics[width=14cm]{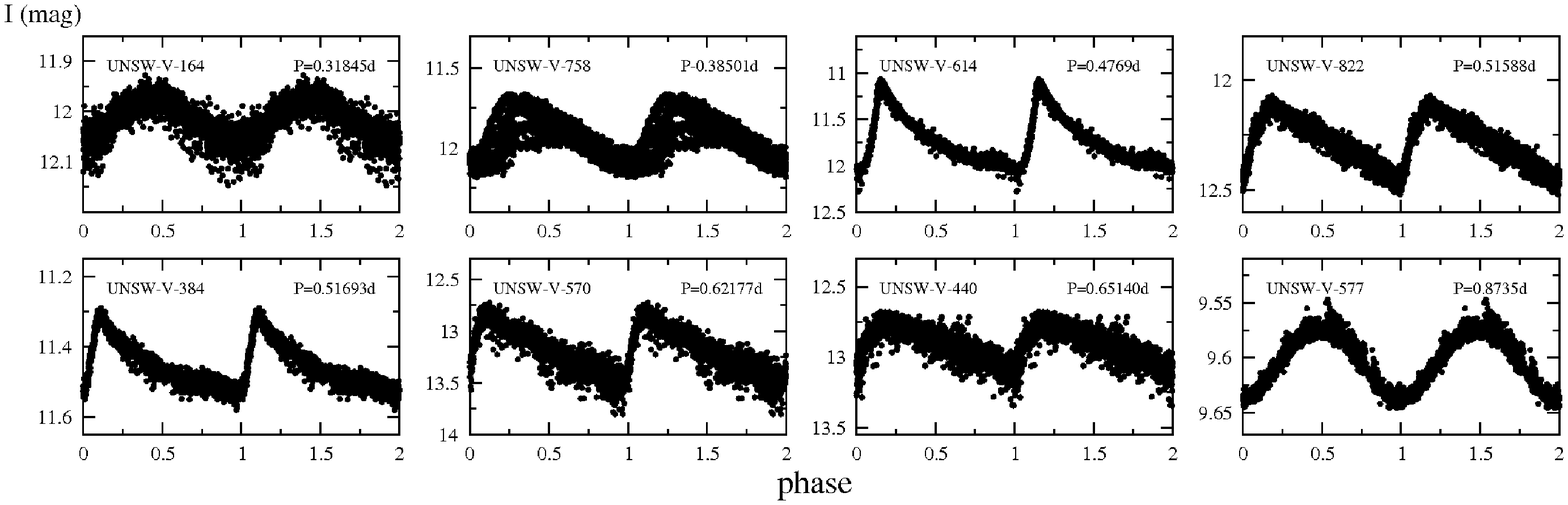}
\includegraphics[width=14cm]{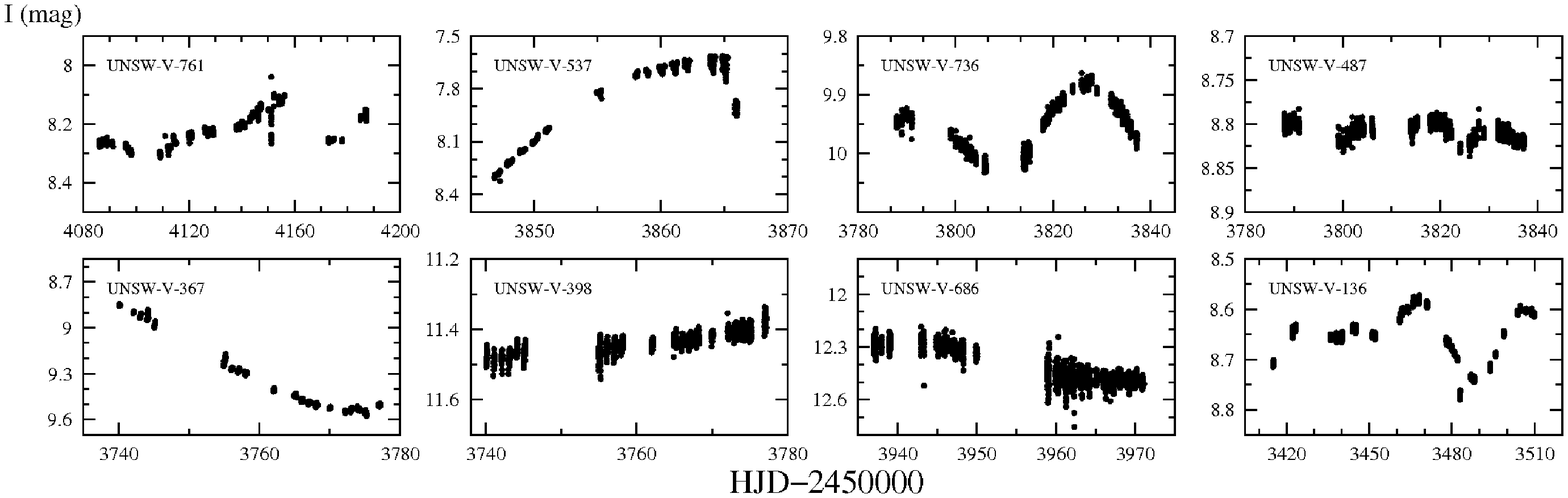}
\caption{\small Sample light curves for contact eclipsing binaries (top three rows), detached
eclipsing binaries (next three rows), RR~Lyrae stars (next two rows) and LPV (bottom two rows). These light curves have not been processed with the trend-filtering algorithm.}
\label{fig:lcsample}
\end{center}
\end{figure*}

As an indication of the quality of the light curves in this catalogue, we plot a representative sample of eclipsing binaries,
pulsating variables and LPVs (Figure \ref{fig:lcsample}). All data are publicly available at the
University of New South Wales Virtual Observatory facility (see Section \ref{sec:online}). Table \ref{tab:full} contains an extract of the complete summary table available in the electronic version of this paper. For each star the ID, J2000 coordinates, galactic coordinates, 2MASS $JHK$ magnitudes, mean $I$-band magnitude, $I$-band amplitude, period, epoch of minimum light, previous identifier where appropriate and classification in this catalogue are shown.

\begin{table*}
 \centering
 \begin{minipage}{\textwidth}
  \caption{Extract from the complete catalogue included in the electronic version of this paper. See text for an explanation of the columns.}
 \label{tab:full}
{\scriptsize 
  \begin{tabular}{@{}lcccccccccccll@{}}
  \hline
  ID       & RA         & Dec         & $l$\dg   & $b$\dg   & $J$    & $H$    & $K$    & $I$   & $A$   & Period  & Epoch     & Alternate ID                 & Type\\
           & (J2000.0)  & (J2000.0)   &          &          & (mag)  & (mag)  & (mag)  & (mag) & (mag) & (d)     & HJD-2450000.0 &                          &     \\
  \hline
UNSW-V-001 & 04:52:56.7 & $-$29:48:14.3 & 231.0476 & $-$37.4715 &  8.258 &  7.805 &  7.650 &  8.66 & 0.012 & -       &       -   &                              & LPV \\
UNSW-V-002 & 04:53:38.1 & $-$29:06:38.0 & 230.2417 & $-$37.1672 & 11.228 & 10.933 & 10.821 & 11.23 & 0.200 & 0.38555 & 3289.1020 & ASAS 045338-2906.6           & EW  \\
UNSW-V-003 & 04:53:48.2 & $-$29:53:49.2 & 231.2140 & $-$37.3108 & 12.475 & 12.078 & 11.969 & 12.71 & 0.026 & 0.69570 & 3289.0900 &                              & EA  \\
UNSW-V-004 & 04:54:43.8 & $-$29:34:07.0 & 230.8696 & $-$37.0407 & 12.388 & 12.111 & 12.083 & 12.46 & 0.018 & -       & -         &                              & LPV \\
UNSW-V-005 & 04:57:28.8 & $-$29:09:48.3 & 230.5523 & $-$36.3636 &  8.540 &  8.235 &  8.160 &  8.75 & 0.005 & 3.2684  & 3289.1200 &                              & EB  \\
UNSW-V-006 & 04:57:45.4 & $-$30:14:05.8 & 231.8658 & $-$36.5529 &  9.342 &  9.162 &  9.109 &  9.38 & 0.003 & -       &       -   &                              & LPV \\
UNSW-V-007 & 04:58:03.5 & $-$29:55:59.1 & 231.5185 & $-$36.4208 & 10.068 &  9.823 &  9.745 & 10.21 & 0.128 & 3.0683  & 3324.2200 & ASAS 045804-2956.0           & EA  \\
UNSW-V-008 & 04:58:18.2 & $-$29:04:54.7 & 230.5073 & $-$36.1695 & 13.720 & 13.279 & 13.238 & 13.73 & 0.050 & 0.31036 & 3288.9700 &                              & EW  \\
UNSW-V-009 & 04:50:06.9 & $-$30:39:50.7 & 231.9448 & $-$38.2536 & 13.548 & 13.424 & 13.386 & 13.69 & 0.139 & 1.0641  & 3351.9800 &                              & EA  \\
UNSW-V-010 & 04:50:18.7 & $-$30:21:29.1 & 231.5754 & $-$38.1480 & 11.697 & 11.139 & 10.976 & 12.31 & 0.032 & 1.9175  & 3290.4200 &                              & PUL \\
  \hline
\end{tabular}
}
\end{minipage}
\end{table*}

\section{Discussion}
\label{sec:dis}

An extensive collection of variable stars always leads to some unexpected results: in the course of analysing transit candidates, the University of New South Wales Extrasolar Planet Search has identified a low-mass K7~Ve detached eclipsing binary (M$_{\rm tot}=1.04\pm0.06$ M$_\odot$,  Young et al. 2006) and the first high-amplitude \dsc~star in an eclipsing binary system (Christiansen et al. 2007). While these alone are interesting, the full breadth of the data is much more extensive. Below we discuss several possible applications, making no attempt at completeness.

\subsection{Close eclipsing binaries with extreme properties}

Contact binaries (or W~UMa-type eclipsing variables) are among the most common types of variable stars,
occurring at a rate of roughly 1 in every 500 FGK dwarfs (Rucinski 2006), which explains their large occurrence rate
in variable star catalogues (e.g. \ew~out of \total~in this catalogue). One intriguing problem
related to these stars is that of binary mergers. When the total angular momentum of a
binary system is at a certain critical (minimum) value, a secular tidal instability occurs which
eventually forces the stars to merge into a single, rapidly rotating object (Arbutina 2007 and
references therein). In the case of contact binaries, the instability occurs at a minimum mass-ratio
of $q_{\rm min}\sim0.071-0.076$ (Rasio 1995, Li \& Zhang 2006), which has been the explanation
for the very few contact systems with $q<0.1$ (see Arbutina 2007 for the updated lists of ten
contact systems in the range of 0.065-0.13). The exact limit depends on assumptions on the
stellar structure and dynamical stability (Li \& Zhang 2006). Since it is likely that at least a
fraction of blue straggler stars in star clusters formed via binary mergers (Mapelli et al. 2004),
there is an exciting opportunity to constrain binary merger theories by increasing the number of known contact binaries with extremely low mass-ratios, and probing the limits of the observed $q_{\rm min}$.

\begin{figure}
\begin{center}
\includegraphics[width=5cm]{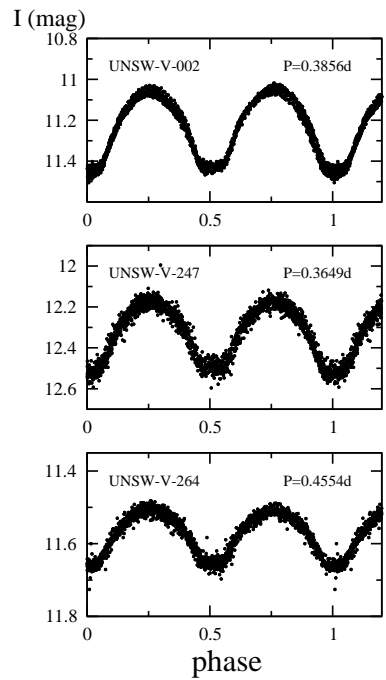}
\caption{\small Possible candidates for low mass-ratio contact binaries.}
\label{fig:qminew}
\end{center}
\end{figure}

Examining the published light curves of the lowest mass-ratio systems (e.g. AW~UMa: Pribulla et
al. 1999; V870~Ara: Szalai et al. 2007), a single flat-bottomed minimum is always present, which
corresponds to the full eclipse of the much smaller component that occurs within a certain range of
inclinations. In our sample we find about five binaries with very similar periods (0.3--0.4~d)
and light curve shapes (three are shown Figure \ref{fig:qminew}), which might therefore be low
mass-ratio systems deserving further attention. This could include obtaining and modelling multi-colour
light curves in several bands (see, e.g., Qian et al. 2005).

Similarly to the mass-ratio, contact binary periods also have a very well-defined cut-off,
which occurs at $P\approx0.215-0.22$~d, just 0.05~d shorter than the maximum of the
volume-corrected period distribution (Rucinski 2007). Stepien (2006) attempted to explain
the period cut-off via the magnetic-wind driven angular momentum loss, the rate of which shows a
progressive decay with the shortening of the period so that the period evolution takes
progressively longer time. The period cut-off would then be due to a finite age of the
binary population of several Gyr. Using the ASAS sample of binaries, Rucinski (2007)
concluded that while no evidence exists for angular momentum evolution, the drop in
numbers towards the cut-off still suffers from small number statistics and the cut-off
itself remains unexplained. Hence, improving the statistics at the short-period end of
contact binaries is important, where high-cadence transit search programs could play an
important role. In our sample, there are four contact binaries out of \ew~in the range of
$P=0.246-0.250$~d, which fall on the short-period end of the distribution but do not improve the
statistics near the cut-off (the present record holder in the Galactic field has a period of
0.2178 d; Rucinski 2007).

It is also possible to use the periods and light curve morphologies to identify close eclipsing binaries that are potentially composed of low-mass components. Identifying low-mass stars in eclipsing binaries is extremely important for accurately deriving the fundamental stellar parameters of mass and radius that are crucial for constraining low-mass stellar formation and evolution models. Following the method of \cite{Weldrake07}, we select those contact binaries with periods $<0.25$~d, and Algol-type detached eclipsing binaries with no obvious out-of-eclipse variations and periods $<1.6$~d as good candidates for low-mass eclipsing binary systems. We find four contact binaries (the same four located near the period cut-off) and 31 detached binaries matching these criteria, listed in Table \ref{tab:lowmass}. There are quite a few bright ($I>12$) objects in this list which would make excellent targets for spectroscopic follow up. 

\begin{table}
\begin{center}
\caption{Eclipsing binary systems potentially composed of low-mass components.}
\label{tab:lowmass}
\begin{tabular}{llllll}
\hline
ID & RA(J2000.0) & Dec(J2000.0) & $I$ (mag) & P (d)\\
\hline
\multicolumn{3}{l}{\bf Contact eclipsing binaries}  &    & \\
UNSW-V-077 & 09 30 05.4 & $-$14 02 17.9 & 12.64 & 0.2480 \\ 
UNSW-V-219 & 16 55 48.1 & $-$60 19 08.7 &13.50  & 0.2456\\ 
UNSW-V-659 & 21 07 03.6 & $-$65 56 42.0 &11.08  & 0.2472\\ 
UNSW-V-662 & 21 10 21.6 & $-$66 54 53.3 &12.37  & 0.2492\\ 
\multicolumn{3}{l}{\bf Detached eclipsing binaries} &     & \\
UNSW-V-003 & 04 53 48.2 & $-$29 53 49.2 & 12.71 & 0.6956 \\ 
UNSW-V-090 & 11 56 40.6 & $-$35 43 43.8 &12.53  & 1.1870\\ 
UNSW-V-097 & 11 59 53.9 & $-$36 13 26.1 &12.18  & 0.9025\\ 
UNSW-V-143 & 16 57 00.6 & $-$60 29 51.9 &12.17 &  0.8117\\
UNSW-V-156 & 17 00 06.2 & $-$60 17 02.4 &13.43  & 1.5618\\ 
UNSW-V-192 & 17 09 18.1 & $-$60 01 43.5 &13.45  & 1.4877\\
UNSW-V-198 & 17 10 18.2 & $-$59 46 08.9 &13.20 &  1.2275\\
UNSW-V-205 & 17 13 55.3 & $-$60 12 15.3 &11.83  & 0.9610\\ 
UNSW-V-301 & 17 16 37.3 & $-$58 09 40.4 &11.04  & 1.4828\\ 
UNSW-V-312 & 00 00 06.0 & $-$59 44 48.3 & 13.63 & 1.0574 \\ 
UNSW-V-353 & 04 04 51.2 & $-$24 11 30.4 & 13.73 & 0.7065 \\ 
UNSW-V-379 & 09 22 49.7 & $-$25 12 40.8 & 13.35 & 0.4975 \\ 
UNSW-V-386 & 09 14 14.5 & $-$24 53 31.6 & 11.43 & 1.2956 \\ 
UNSW-V-410 & 09 13 47.8 & $-$22 48 23.5 & 13.07 & 0.9690 \\ 
UNSW-V-472 & 13 08 49.2 & $-$44 47 55.2 &12.46  & 0.5484\\ 
UNSW-V-527 & 14 45 10.0 & $-$39 25 47.3 &10.99  & 1.363 \\ 
UNSW-V-528 & 14 45 19.4 & $-$38 08 48.0 &11.26  & 1.211 \\ 
UNSW-V-536 & 14 47 27.5 & $-$38 31 36.7 &11.56  & 0.2303\\ 
UNSW-V-540 & 14 49 09.1 & $-$38 38 10.1 &12.92  & 0.5216\\ 
UNSW-V-598 & 18 21 14.1 & $-$64 30 03.1 &11.77  & 0.8492\\ 
UNSW-V-617 & 18 38 00.5 & $-$65 06 07.7 &10.66  & 1.5092\\ 
UNSW-V-621 & 18 11 49.9 & $-$67 06 13.8 &13.76  & 1.0194\\ 
UNSW-V-624 & 18 18 48.4 & $-$67 27 54.1 &12.76  & 1.482 \\ 
UNSW-V-644 & 18 34 50.6 & $-$67 24 09.9 &12.96  & 1.4238\\ 
UNSW-V-683 & 20 58 55.7 & $-$67 02 12.8 &13.21  & 1.0740\\ 
UNSW-V-696 & 21 08 04.4 & $-$68 57 53.1 &11.55  & 0.9428\\ 
UNSW-V-706 & 20 58 35.9 & $-$69 04 01.4 &12.03  & 1.025 \\ 
UNSW-V-722 & 23 25 23.1 & $-$70 01 48.5 &12.20  & 1.3523\\ 
UNSW-V-740 & 08 04 02.0 & $-$66 28 02.6 &13.63  & 1.2634\\
UNSW-V-746 & 08 08 27.0 & $-$68 19 07.5 & 12.68 & 1.0324 \\ 
UNSW-V-846 & 15 03 44.5 & $-$68 40 02.6 &12.13  & 1.5804\\ 
\hline
\end{tabular}
\end{center}
\end{table}

\subsection{Pre-main-sequence eclipsing binaries}

Detached eclipsing binaries provide one of the most accurate (largely model-independent) sources of fundamental stellar parameters, notably masses and radii. These can be used to put the strongest constraints on stellar evolutionary models, which in turn can improve our
understanding of the formation and evolution of individual stellar populations. On the
pre-main-sequence, the calibration of stellar parameters is presently extremely scarce below 1~M$_\odot$ where only six eclipsing binaries are known, all located in the Taurus-Orion region  (Irwin
et al. 2007). Comparison of these systems to different stellar models have indicated difficulties
in fitting both components of the binaries simultaneously, which shows our current models of
low-mass stars are seriously challenged by the known systems (see also \cite{Aigrain07}). 

\begin{figure}
\begin{center}
\includegraphics[width=8.5cm]{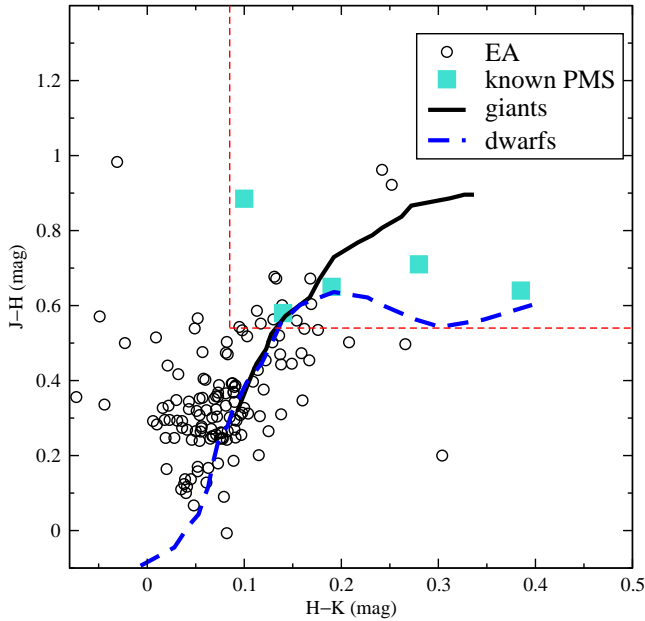}
\caption{\small Detached binaries and the location of five known pre-main-sequence eclipsing 
systems (data taken from Stassun et al. 2004, Covino et al. 2004, Hebb et al. 2006, Stassun,
Mathieu \& Valenti 2007 and Irwin et al. 2007).}
\label{fig:pms2col}
\end{center}
\end{figure}

\begin{table}
\begin{center}
\caption{Candidate PMS detached binaries}
\label{tab:pmscand}
\begin{tabular}{llllll}
\hline
ID & RA(J2000.0) & Dec(J2000.0) & $I$ (mag) & P (d)\\
\hline
UNSW-V-097 & 11 59 53.9 & $-$36 13 26.1 & 12.18 & 0.90251  \\
UNSW-V-107 & 12 03 06.8 & $-$35 37 47.5 & 12.40 & 6.23034  \\
UNSW-V-299 & 17 16 10.8 & $-$57 20 37.5 & 12.33 & 5.97542 \\
UNSW-V-312 & 00 00 06.0 & $-$59 44 48.3 & 13.63 & 1.05762 \\
UNSW-V-491 & 13 12 31.5 & $-$46 04 08.8 & 12.76 & 3.83246 \\
UNSW-V-518 & 13 07 31.8 & $-$45 12 42.0 & 12.59 & 2.75743 \\
UNSW-V-617 & 18 38 00.5 & $-$65 06 07.7 & 10.66 & 1.50906 \\
UNSW-V-676 & 21 20 45.9 & $-$66 22 18.5 & 13.50 & 3.28503 \\
UNSW-V-722 & 23 25 23.1 & $-$70 01 48.5 & 12.20 & 1.01274 \\
UNSW-V-746 & 08 08 27.0 & $-$68 19 07.5 & 12.68 & 1.03214 \\
\hline
\end{tabular}
\end{center}
\end{table}

One possibility for identifying pre-main-sequence binaries is by using colour-colour diagrams,
such as the one depicted in Figure \ref{fig:pms2col}. Here we plot the location of the detached
binaries in our sample, using 2MASS $JHK$ magnitudes, the intrinsic stellar loci from Bessell \&
Brett (1988) and the position of five PMS binaries with published $JHK$ photometry (mostly from
2MASS). We use two dashed lines as boundaries for defining a PMS candidate: the vertical and
horizontal lines are at $H-K=0.085$~mag and $J-H=0.54$~mag, respectively. In total we extract
17 Algol-type systems from our sample that have redder colours than the boundary lines (i.e.
located around the known PMS systems in Figure \ref{fig:pms2col}). Investigation of higher spatial resolution archived images from the
Digitized Sky Surveys\footnote{The Digitized Sky Surveys were produced at the Space Telescope Science Institute under U.S. Government
grant NAG W-2166. The images of these surveys are based on photographic data obtained using the Oschin Schmidt Telescope on Palomar Mountain
and the UK Schmidt Telescope.} (DSS) demonstrates that in six cases, the APT photometry aperture contains a single central 2MASS source.
In four additional cases, the aperture contains a single bright source and up to half a dozen additional sources several magnitudes fainter,
where the depth of the eclipse ($>0.5$~mag) precludes the fainter sources from producing binary signal.

These ten are listed in Table \ref{tab:pmscand}: some could be heavily reddened main-sequence stars, but since the majority of our fields are
located above a Galactic altitude of 15\dg, there is the distinct possibility of genuine PMS binaries in the sample. There is also the
possibility of composite colours of unresolved blends in the 2MASS catalogue. One method of confirmation would be obtaining high-resolution
spectroscopy to look for PMS signatures, such as strong Li absorption (Irwin et al. 2007)

\subsection{The Blazhko-effect in RR~Lyrae stars}

RR~Lyrae stars are horizontal branch stars showing high-amplitude pulsations driven by the
$\kappa$-mechanism, with typical periods of about 0.5 days. It is known that a large fraction of
RR~Lyrae stars (20--30\% of the fundamental mode RRabs and 2\% of the first-overtone RRcs,
Kov\'acs 2001) exhibit periodic amplitude and/or phase modulations, the so-called
Blazhko-effect, which is one the greatest mysteries in classical pulsating star research. Currently, two classes of models are usually put forward as possible explanations, both assuming the
presence of non-radial oscillations (note, that RR~Lyrae stars have long been considered as the
prototypes of purely radially pulsating stars): the resonance models, in which resonance effects
excite non-radial modes in addition to the main radial mode, and the magnetic models, which are
essentially oblique rotating pulsator models (see Kolenberg et al. 2006 and references for more
details). Recently, Stothers (2006) published a new explanation, in which turbulent convection
inside the hydrogen and helium ionization zones becomes cyclically weakened and strengthened
owing to the presence of a transient magnetic field that is generated by some kind of a dynamo
mechanism. 

\begin{table}
\begin{center}
\caption{RR~Lyrae stars with detected Blazhko-effect and double-mode pulsation.}
\label{tab:rrbl}
\begin{tabular}{llllll}
\hline
ID & RA(J2000.0) & Dec(J2000.0) & $I$ (mag) & P (d)\\
\hline
{\bf Blazhko-effect} &      &   &    & \\
UNSW-V-101     & 12 00 49.6 & $-$36 11 59.2 &  14.26 &  0.6264 \\
UNSW-V-203      & 17 12 35.0 & $-$60 29 32.1 &  11.99 &  0.4933 \\
UNSW-V-384 & 09 24 25.5 & $-$24 05 03.4 &  11.45 &  0.5169 \\ 
UNSW-V-442 & 12 50 44.3 & $-$44 41 20.3 &  13.08 &  0.5853 \\ 
UNSW-V-614 & 18 34 41.2 & $-$65 27 08.1 &  11.74 &  0.4769 \\ 
UNSW-V-773  & 14 40 40.8 & $-$68 23 16.8 &  11.03 &  0.5518 \\ 
{\bf Double-mode}  &        &   &     & \\
UNSW-V-358 & 09 16 04.3 &-23 36 08.0& 13.49& 0.3602 \\
           &            &           &      & 0.4840 \\ 
UNSW-V-532 & 14 46 35.8 & -39 33 31.7 &12.86 &0.6540 \\
           &            &             &      & 0.5012\\
UNSW-V-577 & 14 52 42.1 &-41 41 55.3& 9.60 & 0.8735 \\
           &            &           &      & 0.6682\\                                
UNSW-V-758 & 08 16 09.3 &-66 44 46.3& 11.94& 0.3850 \\
           &            &           &      & 0.5170 \\
UNSW-V-810 & 14 50 13.3 &-69 45 18.5& 11.70& 0.7372 \\ 
           &            &           &      & 1.0576\\
\hline
\end{tabular}
\end{center}
\end{table}

With a variety of competing models, theory is in desperate need for further empirical
constraints, most notably ones that are capable of detecting non-radial oscillations and/or magnetic fields, for example high-resolution spectroscopy. Hence, the discovery of bright to moderately faint RR~Lyrae stars with well-expressed Blazhko-effect could be of great interest. In our sample we find six RR~Lyrae stars out
of 52, listed in Table \ref{tab:rrbl}, that demonstrate the Blazhko-effect. Four were previously known variables, and two are new discoveries.
The Blazhko-period of modulation for RR~Lyrae stars typically ranges from tens to hundreds of days (see, for example, fig. 4 of
\cite{Szczygiel07}), although they can be as short as a few days \citep{Jurcsik06}. Due to the comparable baseline of our observations, we
could not determine the Blazhko-period for any of the stars, but the four objects brighter than $I\sim12$ mag, one of which is a new
discovery, are good candidates for further studies.
We also find five double-mode RR~Lyrae stars, for which the period ratios suggest the well-known
mixture of fundamental+first radial overtone pulsation. Three of these are new discoveries, with two previous published in the ASAS-3 catalogue, thus raising the total number of field double-mode RR~Lyrae stars known in the Galaxy to
30; see Szczygiel \& Fabrycky (2007).

\section{Online access to light curves}
\label{sec:online}

All APT images from 2002 July until present day, including those used in the creation of this catalogue, are stored in a publicly available archive. This archive can be accessed using a web browser and the conventional web interface located at: 

\noindent \url{http://astro.ac3.edu.au}

Alternatively, the archive can be accessed via the Simple Image Access Protocol (SIAP) (Tody \& Plante 2004) as defined by the International Virtual Observatory Alliance (IVOA). The SIAP defines a standard for retrieving images from a repository using simple URL-based queries. For example, a request made to the following URL will return a list of APT images that intersect with the 1 degree square region centred on (75,-30) with RA and Dec expressed in decimal degrees:

\noindent \url{http://astro.ac3.edu.au/unsw/siap?POS=75,-30&SIZE=1&TEL=APT}

The POS parameter is mandatory and defines the centre of the search region. The SIZE parameter is optional (default is SIZE=1) and determines the size of the search region. The TEL parameter distinguishes between images from different telescopes, and is specific to the UNSW implementation of the SIAP service.

The list of images returned by the SIAP query is in VOTable format (Ochsenbein et al. 2004). Each item in the list contains a set of attributes describing a particular image that satisfies the search criteria. Also included is a URL which can be used to download the associated image.

The catalogue of light curves discussed in this paper is available from the UNSW archive via the Simple Spectral Access Protocol (SSAP) (Tody et al 2007). The SSAP is an IVOA standard for accessing archives of one dimensional spectra, including time series data such as light curves.  The format of an SSAP query is very similar to the format of an SIAP query.  For example, the query specified in the following URL will search for light curves of stars within a circle of diameter 1 degree centred at the point (75,-30).

\noindent \url{http://astro.ac3.edu.au/unsw/ssap?REQUEST=queryData&POS=75,-30&SIZE=1}

The REQUEST parameter is the only mandatory parameter.  The optional parameters include POS, SIZE, BAND and TIME, which are used to constrain the search by region (degrees), bandpass (metres) and time of observation (ISO 8601).

The list of light curves returned by the SSAP query is in VOTable format, and each item in the list contains a set of metadata describing a particular light curve, including a URL for downloading the data.  The light curve data itself is also presented in VOTable format and follows the structure of the Spectrum Data Model (McDowell 2007).  Consequently, these light curve VOTables may be examined with any VO-compliant tool, such as TOPCAT (Taylor 2005).

\section{Summary}
\label{sec:summary}

We have presented a catalogue of 850 variable stars, compiled from 32 months of observations obtained for the University of New South Wales Extrasolar Planet Search. Of these stars, 659 are new discoveries that have not been previously reported in the GCVS, NSV or ASAS-3 catalogues. This catalogue of well-sampled high precision light curves, each spanning 1--4~months, has significant potential for astrophysically interesting data mining. We have nominated several possibilities, including eclipsing binary systems with low mass ratios, low-mass components or pre-main sequence components, and RR Lyrae stars demonstrating the curious Blazhko-effect. The data have been made publicly available on the University of New South Wales VO server in a standard format for retrieval and for analysis with standard VO tools.

\section*{Acknowledgments}

We thank the referee for helpful comments. This project has been supported by the Australian Research Council and the Australian Research Collaboration Service (ARCS). JLC and AD are supported by Australian Postgraduate Research Awards.

\bsp

\label{lastpage}

\appendix

\section{Cross-identifications with the GCVS, ASAS and ROSAT catalogues}

The positions of the stars in this catalogue were correlated with the General Catalog of
Variable Stars (GCVS) and All Sky Automated Survey (ASAS) variable star catalogues, using the
{\em Vizier} online database \citep{Ochsenbein00}. The photometry aperture used in our data
reduction pipeline has a radius of 28.2$^{\prime\prime}$, and so a simple cone search with a
radius of 30$^{\prime\prime}$ was performed. The results of the correlation are shown in Table
\ref{tab:gcvs}. 191 of the 850 stars presented in this catalogue are positionally
coincident with previously published variable stars. The columns in the table are the UNSW
identifier from this catalogue, the right ascension and declination (J2000.0), the mean $I$-band
magnitude, $I$-band amplitude of variation, period, epoch, the identifier from either the GCVS, NSV or
ASAS catalogues, and our classification, which was found to be in good agreement with the published
classification in more than 90\% of the cases with a few exceptions, like V717~Ara (EB), which is 
listed as an RR~Lyr in the GCVS or V500~Ara (EW), also RR~Lyr in the GCVS. However, these are the
classes with highly sinusoidal, i.e. indistinguishable light curve shapes, and it is therefore not
surprising that single-filtered light curves are not enough in doubtful cases.

The positions were also correlated with x-ray sources in the ROSAT 1RXS \citep{Voges99,Voges00} and
2RXP (ROSAT 2000) catalogues, similarly to \cite{Norton07}. The search was again performed through
{\em Vizier} using a 30$^{\prime\prime}$ cone search, and the results are shown in Table
\ref{tab:rosat}. The columns are as for Table \ref{tab:gcvs}1, although in this case the second
identifier column contains the ROSAT source identifier. 22 of the 850 stars were found to be
spatially coincident with ROSAT sources, although we note that since the majority have been
classified as pulsating variables, which are not expected to be strong X-ray sources (Makarov 2003), it is
doubtful how much of the coincidence is real.

\begin{table*}
 \centering
 \begin{minipage}{140mm}
  \caption{UNSW variable stars coincident with GCVS/ASAS records.}
 \label{tab:gcvs}
{\scriptsize 
  \begin{tabular}{@{}lccccccll@{}}
  \hline
  ID & RA         & Dec        & $I$ & $A$ & Period & Epoch & Alternate ID & Type\\
     & (J2000.0)  & (J2000.0)  & (mag) & (mag) & (d) & HJD-2450000.0 & & \\
  \hline
UNSW-V-002  &   04 53 38.1 & $-$29 06 38.0 &	11.23 & 0.200 &   0.38555 &  3289.1020 &  ASAS 045338-2906.6	      &   EW		   \\
UNSW-V-007  &   04 58 03.5 & $-$29 55 59.1 &	10.21 & 0.128 &   3.06827 &  3324.2200 &  ASAS 045804-2956.0	      &   EA		   \\
UNSW-V-011  &   05 00 55.6 & $-$30 07 54.1 &	12.06 & 0.127 &   0.84183 &  3291.1200 &  ASAS 050056-3007.9	      &   EW		   \\
UNSW-V-013  &   05 02 46.6 & $-$29 44 04.2 &	 8.41 & 0.026 &   -	  &  -     &  ASAS 050247-2944.1	      &   LPV		   \\
UNSW-V-016  &   04 51 51.6 & $-$29 34 22.4 &	11.65 & 0.209 &   0.38150 &  3289.3880 &  ASAS 045151-2934.3	      &   EW		   \\
UNSW-V-018  &   04 42 00.0 & $-$25 28 45.5 &	13.52 & 0.134 &   0.58747 &  3289.1000 &  ASAS 044200-2528.8	      &   RRL		   \\
UNSW-V-019  &   04 42 16.3 & $-$25 49 32.1 &	11.46 & 0.220 &   0.25489 &  3285.3350 &  ASAS 044216-2549.5	      &   EW		   \\
UNSW-V-021  &   04 42 25.4 & $-$26 17 31.3 &	11.52 & 0.087 &   0.45552 &  3285.2000 &  ASAS 044225-2617.5	      &   EW		   \\
UNSW-V-026  &   04 48 00.9 & $-$25 31 23.1 &	12.23 & 0.200 &   0.54883 &  3285.1700 &  ASAS 044801-2531.4	      &   RRL		   \\
UNSW-V-035  &   09 03 32.6 & $-$14 52 45.5 &	11.12 & 0.070 &   0.29358 &  3376.9850 &  ASAS 090333-1452.8		      &   EW		   \\
UNSW-V-043  &   09 05 55.6 & $-$14 51 23.4 &	10.90 & 0.069 &   0.56636 &  3377.1500 &  ASAS 090556-1451.4		      &   EW		   \\
UNSW-V-046  &   09 07 07.5 & $-$14 07 37.7 &	 9.84 & 0.041 &   -	  &  -     &  ASAS 090708-1407.7	      &   LPV		   \\
UNSW-V-050  &   09 08 20.6 & $-$13 35 43.2 &	11.14 & 0.068 &  15.44279 &  3389.0400 &  ASAS 090821-1335.8		  &   EA	 \\
UNSW-V-055  &   08 59 54.2 & $-$15 15 22.9 &	11.57 & 0.186 &   0.39362 &  3371.1350 &  ASAS 085954-1515.4		      &   EW		   \\
UNSW-V-057  &   09 22 21.2 & $-$13 38 50.4 &	12.50 & 0.188 &   0.53220 &  3377.9800 &  IV HYA/ASAS 092220-1338.8	      &   RRL		   \\
UNSW-V-061  &   09 23 55.7 & $-$14 20 22.1 &	 8.65 & 0.026 &  35.06803 &  3393.0000 &  ASAS 092356-1420.4		      &   LPV		   \\
UNSW-V-062  &   09 24 55.9 & $-$13 11 58.2 &	12.92 & 0.180 &   0.61124 &  3382.1500 &  ASAS 092456-1312.0		      &   RRL		   \\
UNSW-V-066  &   09 26 41.0 & $-$13 45 06.5 &	 9.66 & 0.239 &   0.44976 &  3377.2500 &  EZ HYA/ASAS 092641-1345.1	  &	  EW		   \\
UNSW-V-070  &   09 28 20.1 & $-$12 50 51.2 &	11.70 & 0.248 &   0.50182 &  3377.0970 &  ASAS 092820-1250.9		      &   EB		   \\
UNSW-V-071  &   09 28 28.7 & $-$13 26 33.2 &	11.68 & 0.205 &   0.73146 &  3378.4340 &  ASAS 092829-1326.5		      &   EB		   \\
UNSW-V-074  &   09 28 57.5 & $-$13 07 12.4 &	 8.16 & 0.029 &  45.49365 &  3402.0000 &  ASAS 092857-1307.2		      &   LPV		   \\
UNSW-V-075  &   09 29 15.3 & $-$14 05 57.2 &	12.27 & 0.263 &   0.33259 &  3377.2450 &  ASAS 092915-1405.9		      &   EW		  \\
UNSW-V-081  &   09 20 17.4 & $-$14 05 09.7 &	 8.61 & 0.099 &   -	  &  -     &  ASAS 092018-1405.1	      &   LPV		   \\
UNSW-V-084  &   11 55 44.0 & $-$36 26 19.8 &	11.78 & 0.134 &   0.53132 &  3415.1900 &  ASAS 115544-3626.3		      &   EW		   \\
UNSW-V-086  &   11 56 12.9 & $-$35 59 29.9 &	11.50 & 0.266 &   0.37801 &  3415.1180 &  V0576 CEN/ASAS 115613-3559.5    &	  EW		   \\
UNSW-V-087  &   11 56 20.4 & $-$35 28 45.5 &	11.39 & 0.141 &   0.29374 &  3415.3250 &  ASAS 115620-3528.8		      &   EW		   \\
UNSW-V-090  &   11 56 40.6 & $-$35 43 43.8 &	12.53 & 0.102 &   1.18713 &  3415.1300 &  V577 CEN			      &   EA		   \\
UNSW-V-091  &   11 57 20.2 & $-$36 40 23.3 &	13.52 & 0.133 &   0.58688 &  3422.9900 &  V0580 CEN			      &   RRL		   \\
UNSW-V-092  &   11 57 57.2 & $-$36 06 10.3 &	13.41 & 0.266 &   0.34622 &  3415.1670 &  V0581 CEN			      &   EW		   \\
UNSW-V-097  &   11 59 53.9 & $-$36 13 26.1 &	12.18 & 0.057 &   0.90251 &  3422.1700 &  NSV 05410		      &   EA		   \\
UNSW-V-101  &   12 00 49.6 & $-$36 11 59.2 &	14.26 & 0.170 &   0.62644 &  3423.0900 &  V0582 CEN			      &   RRL		   \\
UNSW-V-105  &   12 02 27.6 & $-$35 26 39.3 &	14.04 & 0.143 &   0.67441 &  3423.0800 &  EF HYA		      &   RRL		   \\
UNSW-V-107  &   12 03 06.8 & $-$35 37 47.5 &	12.40 & 0.078 &   6.23034 &  3415.0100 &  NSV 05437		      &   EA	  \\
UNSW-V-115  &   11 53 52.2 & $-$35 26 55.4 &	13.57 & 0.084 &   0.58367 &  3415.1200 &  DS HYA		      &   RRL		   \\
UNSW-V-116  &   11 54 06.9 & $-$35 14 54.8 &	11.22 & 0.082 &   3.30294 &  3423.1900 &  ASAS 115407-3514.9		      &   EA		   \\
UNSW-V-122  &   11 56 27.7 & $-$38 08 51.3 &	 8.85 & 0.077 &  84.16663 &  -         &  ASAS 115628-3808.9	      &   LPV		   \\
UNSW-V-131  &   12 01 25.6 & $-$37 24 52.6 &	 8.35 & 0.583 & 135.44203 &  -         &  V0583 CEN/ASAS 120125-3724.9    &   LPV		   \\
UNSW-V-132  &   12 02 19.3 & $-$38 46 22.3 &	12.86 & 0.264 &   0.45906 &  3415.0900 &  V0584 CEN/ASAS 120219-3846.4    &	  RRL		   \\
UNSW-V-137  &   12 06 53.6 & $-$37 37 36.0 &	12.51 & 0.157 &   0.39816 &  3415.3260 &  ASAS 120654-3737.7		      &   EW		   \\
UNSW-V-138  &   11 53 40.7 & $-$38 22 00.5 &	10.55 & 0.063 &   0.82146 &  3423.5700 &  ASAS 115341-3822.1		      &   EW		   \\
UNSW-V-140  &   16 56 35.0 & $-$59 04 41.2 &	10.44 & 0.580 & 184.16278 &  -         &  CG ARA			      &   LPV		   \\
UNSW-V-143  &   16 57 00.6 & $-$60 29 51.9 &	12.17 & 0.020 &   0.81166 &  3510.1600 &  V0805 ARA			      &   EA		   \\
UNSW-V-144  &   16 57 15.2 & $-$60 08 05.3 &	12.71 & 0.224 &   0.92708 &  3499.5400 &  V0705 ARA			      &   EW		   \\
UNSW-V-145  &   16 57 50.6 & $-$59 07 20.0 &	12.60 & 0.187 &   0.57442 &  3509.1100 &  V0414 ARA			      &   RRL		   \\
UNSW-V-147  &   16 57 43.8 & $-$60 41 16.4 &	 9.33 & 0.086 &   1.20159 &  3499.8030 &  ASAS 165744-6041.3	      &   EW		   \\
UNSW-V-154  &   16 59 07.2 & $-$60 05 25.4 &	 8.55 & 0.903 & 174.95479 &  -         &  LS ARA/ASAS 165907-6005.4       &   LPV		   \\
UNSW-V-171  &   17 02 35.4 & $-$59 30 48.7 &	 9.84 & 0.302 & 118.49562 &  -         &  V0810 ARA 		      &   LPV		   \\
UNSW-V-178  &   16 54 36.1 & $-$60 09 04.8 &	10.68 & 0.375 &   -	  &   -        &  V0696 ARA			      &   LPV		   \\
UNSW-V-184  &   17 05 59.3 & $-$59 40 34.1 &	12.63 & 0.033 &   0.51838 &  3499.1100 &  V0464 ARA			      &   RRL		   \\
UNSW-V-186  &   17 06 48.2 & $-$59 03 20.0 &	 8.68 & 0.864 & 152.13318 &  -         &  CH ARA/ASAS 170648-5903.3       &   LPV		   \\
UNSW-V-187  &   17 07 32.4 & $-$60 58 45.6 &	11.47 & 0.117 &   2.32845 &  3504.1400 &  ASAS 170732-6058.8	      &   EA		   \\
UNSW-V-194  &   17 09 43.0 & $-$60 31 14.9 &	12.27 & 0.170 &   0.35894 &  3499.2330 &  ASAS 170944-6031.2	      &   EW		   \\
UNSW-V-195  &   17 10 07.8 & $-$60 39 45.7 &	10.51 & 0.134 &   2.52195 &  3499.2000 &  V0617 ARA/ASAS 171008-6039.8    &	  CEP		   \\
UNSW-V-198  &   17 10 18.2 & $-$59 46 08.9 &	13.20 & 0.072 &   1.22754 &  3555.2400 &  V0485 ARA			      &   EA		   \\
UNSW-V-199  &   17 11 16.1 & $-$60 20 42.8 &	 8.69 & 0.637 & 184.16278 &  -	       &  CN ARA			      &   LPV		   \\
UNSW-V-200  &   17 11 32.5 & $-$60 14 35.9 &	 9.46 & 0.273 & 152.13318 &  - 	       &  NSV 08245 		      &   LPV		   \\
UNSW-V-203  &   17 12 35.0 & $-$60 29 32.1 &	11.99 & 0.045 &   0.49332 &  3504.2800 &  CS ARA		      &   RRL		   \\
UNSW-V-207  &   17 13 52.6 & $-$59 05 16.1 &	 9.56 & 0.137 &  75.56063 &  -         &  V0733 ARA/ASAS 171352-5905.2    &   LPV		   \\
UNSW-V-214  &   17 15 58.6 & $-$60 04 06.2 &	11.25 & 0.129 &   0.40006 &  3499.1600 &  ASAS 171559-6004.1	      &   EW		   \\
UNSW-V-215  &   17 16 19.3 & $-$59 55 07.2 &	12.77 & 0.072 &   0.77541 &  3504.2500 &  DG ARA		      &   RRL		   \\
UNSW-V-217  &   17 16 35.5 & $-$60 21 01.4 &	11.65 & 0.204 &   0.36364 &  3499.1060 &  V0791 ARA/ASAS 171636-6021.1    &	  EW		   \\
UNSW-V-218  &   17 17 40.0 & $-$60 31 19.1 &	11.77 & 0.066 &   0.59981 &  3499.1400 &  MT ARA		      &   EB		   \\
UNSW-V-229  &   17 01 00.5 & $-$58 35 18.1 &	12.12 & 0.066 &   1.28635 &  3504.2500 &  V717 ARA			      &   EW		   \\
UNSW-V-231  &   17 01 23.8 & $-$58 16 46.1 &	12.90 & 0.036 &   0.58292 &  3504.1000 &  V0441 ARA			      &   RRL		   \\
UNSW-V-232  &   17 01 33.8 & $-$58 18 35.1 &	13.47 & 0.047 &   2.04759 &  3558.1000 &  V0442 ARA			      &   EA		   \\
UNSW-V-233  &   17 01 49.3 & $-$57 59 33.5 &	10.25 & 0.712 & 166.85367 &	   -     &  V0779 ARA 		      &   LPV		   \\
UNSW-V-238  &   17 02 59.7 & $-$57 06 42.8 &	 9.49 & 0.080 &   0.98892 &  3510.2800 &  V0722 ARA			      &   EA		   \\
UNSW-V-248  &   16 57 56.7 & $-$57 52 46.4 &	10.94 & 0.471 &   -	  &  -     &  V0776 ARA		              &   LPV		   \\
UNSW-V-252  &   17 06 37.5 & $-$58 07 40.2 &	11.02 & 0.124 &   0.89702 &  3504.2500 &  ASAS 170637-5807.7	      &   EW		   \\
UNSW-V-254  &   17 06 33.9 & $-$57 42 51.4 &	 8.71 & 0.163 & 129.34859 &	   -     &  NSV 08172/ASAS 170634-5742.9    &   LPV		   \\
UNSW-V-255  &   17 06 46.4 & $-$58 31 15.2 &	12.05 & 1.066 & 175.19339 &	   -     &  V0780 ARA 		      &   LPV		   \\
UNSW-V-259  &   16 58 20.3 & $-$57 19 38.9 &	11.20 & 0.031 &   0.11082 &  3504.2200 &  V0709 ARA			      &   PULS         \\
UNSW-V-262  &   17 09 25.7 & $-$57 44 43.4 &	12.60 & 0.099 &   0.66760 &  3504.5700 &  V0817 ARA			      &   EB		   \\
UNSW-V-264  &   17 09 44.2 & $-$57 53 42.2 &	11.57 & 0.075 &   0.45536 &  3504.0700 &  ASAS 170944-5753.7	      &   EW		   \\
UNSW-V-265  &   17 09 55.7 & $-$58 43 14.5 &	12.32 & 0.014 &   0.90689 &  3504.2800 &  V0783 ARA			      &   EA		   \\
UNSW-V-267  &   17 10 00.8 & $-$58 10 08.9 &	11.82 & 0.154 &   0.41460 &  3504.0800 &  CL ARA/ASAS 171000-5810.2	      &   EW		  \\
UNSW-V-268  &   17 10 01.4 & $-$57 58 26.0 &	 9.87 & 0.017 &   8.56234 &  3581.5000 &  ASAS 171002-5758.4		      &   EA	 \\
UNSW-V-269  &   17 10 15.8 & $-$57 26 47.2 &	10.05 & 0.087 &  58.40231 &	   -     &  V0731 ARA/ASAS 171016-5726.7    &   LPV		  \\
  \hline
\end{tabular}
}
\end{minipage}
\end{table*}

\begin{table*}
 \centering
 \begin{minipage}{140mm}
  \contcaption{}
 \label{tab:gcvs2}
{\scriptsize 
  \begin{tabular}{@{}lcccccclll@{}}
  \hline
  ID & RA         & Dec        & $I$ & $A$ & Period & Epoch & Alternate ID & Type\\
     & (J2000.0)  & (J2000.0)  & (mag) & (mag) & (d) & HJD-2450000.0 & & \\
  \hline
UNSW-V-271  &   17 11 13.0 & $-$57 14 03.1 &	13.04 & 0.230 &   0.96374 &  3504.1900 &  V0785 ARA			      &   EB		   \\
UNSW-V-274  &   17 11 24.2 & $-$57 00 47.6 &	12.30 & 0.129 &   0.49237 &  3503.9700 &  V0492 ARA			      &   EW		   \\
UNSW-V-277  &   17 11 58.9 & $-$57 36 53.2 &	10.39 & 0.918 & 166.85367 &	   -     &  V0493 ARA/ASAS 171200-5737.2    &   LPV		   \\
UNSW-V-278  &   17 12 09.1 & $-$58 34 15.1 &	10.66 & 0.543 & 166.85367 &	   -     &  CQ ARA/ASAS 171209-5834.2       &   LPV		   \\
UNSW-V-280  &   17 12 34.4 & $-$57 24 11.1 &	11.23 & 0.823 & 122.58292 &	   -     &  CT ARA			      &   LPV		   \\
UNSW-V-285  &   17 13 33.5 & $-$57 55 15.3 &	10.57 & 0.389 & 119.22419 &	   -     &  V0732 ARA/ASAS 171334-5755.2    &   LPV		   \\
UNSW-V-287  &   17 14 24.3 & $-$58 46 39.6 &	10.39 & 0.918 & 140.15567 &	   -     &  V0498 ARA 		      &   LPV		   \\
UNSW-V-289  &   17 14 28.1 & $-$57 26 15.9 &	12.97 & 0.177 &   0.39302 &  3504.2300 &  V0500 ARA			      &   EW		   \\
UNSW-V-295  &   17 15 33.2 & $-$57 05 15.1 &	10.46 & 1.149 & 175.19339 &	   -     &  DE ARA			      &   LPV		   \\
UNSW-V-301  &   17 16 37.3 & $-$58 09 40.4 &	11.04 & 0.032 &   1.48306 &  3531.2800 &  ASAS 171638-5809.7	      &   EA		   \\
UNSW-V-307  &   17 18 45.7 & $-$57 46 29.7 &	11.17 & 0.208 &   0.79384 &  3510.2070 &  NSV 08452/ASAS 171846-5746.5    &	  EB		   \\
UNSW-V-308  &   17 18 45.1 & $-$57 26 20.8 &	 8.03 & 0.045 &   1.80572 &  3499.6000 &  V0858 ARA			      &   PUL		   \\
UNSW-V-310  &   23 57 02.7 & $-$58 26 01.3 &	13.42 & 0.223 &   0.68809 &  3582.2300 &  ASAS 235702-5826.0	      &   RRL		   \\
UNSW-V-317  &   00 04 15.9 & $-$58 15 53.5 &	 9.11 & 0.060 & 143.47872 &	   -     &  ASAS 000416-5815.9	      &   LPV		   \\
UNSW-V-327  &   00 01 47.5 & $-$57 14 30.4 &	10.09 & 0.082 &   0.47036 &  3579.1760 &  ASAS 000147-5714.5	      &   EW		   \\
UNSW-V-329  &   00 02 29.3 & $-$56 53 49.9 &	 8.79 & 0.042 &  13.21167 &  3591.5000 &  ASAS 000229-5653.9	      &   CEP		   \\
UNSW-V-336  &   23 51 57.4 & $-$57 25 20.8 &	10.01 & 0.088 &   0.39260 &  3577.1370 &  ASAS 235157-5725.4	      &   EW		   \\
UNSW-V-337  &   23 54 23.7 & $-$57 56 27.6 &	10.40 & 0.199 &   0.58423 &  3577.5100 &  ASAS 235424-5756.5	      &   EW		   \\
UNSW-V-352  &   04 13 38.9 & $-$24 33 32.5 &	12.05 & 0.066 &   -	  &  -     &  ASAS 041339-2433.5	      &   EW		   \\
UNSW-V-367  &   09 18 55.4 & $-$25 16 44.1 &	 9.14 & 0.398 &  59.06429 &	   -     &  Z PYX/ASAS 091855-2516.7        &   LPV		   \\
UNSW-V-369  &   09 19 41.8 & $-$24 18 38.5 &	 9.96 & 0.081 &  36.01291 &	   -     &  ASAS 091942-2418.6	      &   LPV		   \\
UNSW-V-370  &   09 20 00.7 & $-$23 38 42.9 &	 8.51 & 0.034 &  52.60413 &	   -     &  ASAS 092000-2338.7	      &   LPV		   \\
UNSW-V-377  &   09 22 37.7 & $-$25 27 06.4 &	12.18 & 0.212 &   0.48368 &  3740.2300 &  SS PYX/ASAS 092238-2527.1 1	      &   EW		   \\
UNSW-V-384  &   09 24 25.5 & $-$24 05 03.4 &	11.45 & 0.076 &   0.51693 &  3742.0900 &  ASAS 092425-2405.1		      &   RRL		   \\
UNSW-V-390  &   09 25 51.5 & $-$24 00 39.4 &	 7.98 & 0.242 &  59.06429 &	   -     &  LP HYA			      &   LPV		   \\
UNSW-V-396  &   09 10 29.0 & $-$22 44 34.4 &	 8.71 & 0.025 &  35.67026 &	   -     &  ASAS 091029-2244.6	      &   LPV		   \\
UNSW-V-400  &   09 11 03.1 & $-$23 27 16.3 &	11.69 & 0.168 &   0.62330 &  3743.0300 &  ASAS 091103-2327.3		      &   EW		   \\
UNSW-V-421  &   09 17 26.3 & $-$22 48 10.7 &	 9.03 & 0.120 &  59.06429 &	   -     &  ASAS 091726-2248.2	      &   LPV		   \\
UNSW-V-450  &   12 54 31.3 & $-$46 07 36.5 &	 8.68 & 0.014 &   1.04272 &  3805.1500 &  NSV 06020		      &   CEP		   \\
UNSW-V-461  &   12 47 23.7 & $-$45 35 03.3 &	 9.26 & 0.075 &   -	  &  -     &  ASAS 124724-4535.1	      &   LPV		   \\
UNSW-V-473  &   13 08 47.6 & $-$45 56 58.3 &	 8.35 & 0.084 &  67.33070 &	   -     &  ASAS 130848-4557.3	      &   LPV	 \\
UNSW-V-477  &   13 10 17.1 & $-$44 25 59.7 &	 8.08 & 0.025 &  48.60485 &	   -     &  ASAS 131017-4426.2	      &   LPV	 \\
UNSW-V-480  &   13 10 31.5 & $-$45 19 31.4 &	 9.30 & 0.085 &  49.98603 &	   -     &  NSV 06118 		      &   LPV	 \\
UNSW-V-494  &   13 13 20.7 & $-$45 38 13.4 &	 8.21 & 0.115 &  48.60485 &	   -     &  ASAS 131321-4538.5	      &   LPV	 \\
UNSW-V-495  &   13 13 33.0 & $-$44 49 31.8 &	 9.10 & 1.206 &  57.39314 &	   -     &  ASAS 131333-4449.8	      &   LPV		   \\
UNSW-V-500  &   13 10 18.5 & $-$45 08 59.8 &	11.38 & 0.031 &   5.35927 &  3787.9950 &  ASAS 131018-4509.2		      &   EA+DSCT	   \\
UNSW-V-503  &   13 15 39.9 & $-$45 51 14.9 &	12.18 & 0.151 &   0.33926 &  3788.1100 &  ASAS 131540-4551.5		      &   EW		   \\
UNSW-V-508  &   13 16 49.0 & $-$45 48 47.2 &	 8.82 & 0.195 &   0.40969 &  3788.1050 &  ASAS 131649-4549.0	      &   EW		   \\
UNSW-V-516  &   13 18 21.8 & $-$44 43 28.1 &	12.08 & 0.074 &   0.30021 &  3788.0000 &  ASAS 131823-4443.9		      &   RRL		   \\
UNSW-V-520  &   13 21 15.1 & $-$45 13 21.9 &	 8.36 & 0.018 &  29.35691 &  4000.0000 &  ASAS 132115-4513.6		      &   LPV		   \\
UNSW-V-524  &   14 43 52.6 & $-$39 54 40.2 &	10.07 & 0.265 &  31.46413 &	   -     &  V0549 CEN 		      &   LPV		   \\
UNSW-V-535  &   14 47 23.2 & $-$39 06 22.6 &	11.80 & 0.158 &   0.31378 &  3846.9350 &  ASAS 144723-3906.4	      &   EW		   \\
UNSW-V-537  &   14 48 14.5 & $-$38 18 32.7 &	 8.44 & 0.795 &  29.14202 &	   -     &  V0557 CEN/ASAS 144815-3818.6    &   LPV		   \\
UNSW-V-551  &   14 52 30.1 & $-$38 24 33.2 &	10.44 & 0.082 &  30.28946 &	   -     &  ASAS 145230-3824.6	      &   LPV		   \\
UNSW-V-556  &   14 54 44.7 & $-$38 56 47.8 &	 9.81 & 0.135 &  31.77751 &	   -     &  V0566 CEN 		      &   LPV		   \\
UNSW-V-557  &   14 42 20.5 & $-$38 40 32.9 &	12.76 & 0.103 &   3.09517 &  3848.9600 &  V0544 CEN			      &   EA	  \\
UNSW-V-558  &   14 43 27.3 & $-$41 02 05.7 &	 8.07 & 0.050 &   -	  &  -     &  V0642 CEN 		      &   LPV		   \\
UNSW-V-562  &   14 45 49.0 & $-$41 26 12.4 &	 8.51 & 0.013 &   -	  &  -     &  V0551 CEN 		      &   LPV		   \\
UNSW-V-564  &   14 46 58.5 & $-$41 17 50.6 &	12.82 & 0.083 &   0.69187 &  3847.2000 &  NSV 06795			      &   RRL		   \\
UNSW-V-567  &   14 48 05.4 & $-$41 45 23.6 &	10.47 & 0.186 &   -	  &  -     &  V0555 CEN/ASAS 144805-4145.4    &   LPV		   \\
UNSW-V-570  &   14 49 00.4 & $-$41 26 55.0 &	13.17 & 0.226 &   0.62177 &  3847.1400 &  V0558 CEN			      &   RRL		   \\
UNSW-V-571  &   14 49 31.6 & $-$40 06 29.1 &	 9.01 & 0.074 &   0.79930 &  3847.1250 &  ASAS 144932-4006.4	      &   EW		   \\
UNSW-V-572  &   14 49 24.8 & $-$40 04 27.9 &	 9.14 & 0.042 &   -	  &  -     &  V0560 CEN 		      &   LPV		   \\
UNSW-V-574  &   14 50 28.2 & $-$40 56 15.0 &	11.99 & 0.208 &   0.47623 &  3847.1100 &  ASAS 145028-4056.3	      &   EW		   \\
UNSW-V-577  &   14 52 42.1 & $-$41 41 55.3 &	 9.60 & 0.032 &   0.87351 &  3847.0400 &  ASAS 145242-4141.9		      &   RRL  \\
UNSW-V-583  &   14 42 34.7 & $-$40 27 17.4 &	10.00 & 0.184 &   0.32503 &  3846.9750 &  V0677 CEN/ASAS 144235-4027.2    &	  EW		   \\
UNSW-V-584  &   14 42 55.3 & $-$41 18 47.4 &	 9.70 & 0.088 &   -	  &  -     &  V0545 CEN 		      &   LPV		   \\
UNSW-V-591  &   18 18 43.6 & $-$64 37 59.6 &	 9.55 & 0.043 &   -	  &  -     &  ASAS 181843-6437.9	      &   LPV		   \\
UNSW-V-592  &   18 19 04.5 & $-$65 35 35.3 &	 9.18 & 0.473 &   -	  &  -     &  DF PAV/ASAS 181904-6535.6       &   LPV		   \\
UNSW-V-595  &   18 20 32.2 & $-$64 17 23.7 &	 8.51 & 0.033 &   -	  &  -     &  ASAS 182031-6417.3	      &   LPV		   \\
UNSW-V-599  &   18 21 32.3 & $-$64 15 58.1 &	 8.59 & 0.416 &   -	  &  -     &  NSV 10616 		      &   LPV		   \\
UNSW-V-601  &   18 22 36.6 & $-$65 30 18.4 &	 9.63 & 0.076 &   -	  &  -     &  ASAS 182236-6530.3	      &   LPV		   \\
UNSW-V-603  &   18 24 38.7 & $-$65 11 02.0 &	10.49 & 0.059 &   2.41936 &  3879.1400 &  ASAS 182438-6511.0	      &   EA		   \\
UNSW-V-606  &   18 26 23.7 & $-$64 57 45.9 &	12.68 & 0.072 &   2.49162 &  3872.8300 &  DP PAV		      &   EA		   \\
UNSW-V-607  &   18 29 37.0 & $-$64 54 43.1 &	 7.39 & 1.814 &   -	  &  -     &  NSV 10827/ASAS 182937-6454.7    &   LPV		   \\
UNSW-V-609  &   18 13 35.1 & $-$65 14 13.1 &	10.86 & 0.463 &   -	  &  -     &  NW PAV/ASAS 181335-6514.2       &   CEP		   \\
UNSW-V-610  &   18 30 35.7 & $-$64 51 33.7 &	 8.69 & 0.031 &   -	  &  -     &  ASAS 183034-6451.5	      &   LPV		   \\
UNSW-V-614  &   18 34 41.2 & $-$65 27 08.1 &	11.74 & 0.291 &   0.47690 &  3874.1000 &  BH PAV/ASAS 183441-6527.0	      &   RRL	   \\
UNSW-V-615  &   18 35 24.0 & $-$64 57 04.4 &	 9.24 & 0.300 &   -	  &  -     &  ASAS 183523-6457.0	      &   LPV		   \\
UNSW-V-623  &   18 18 32.0 & $-$67 19 48.5 &	 9.21 & 0.078 &   -	  &  -     &  ASAS 181833-6719.9	      &   LPV		   \\
UNSW-V-627  &   18 21 15.4 & $-$66 38 47.7 &	11.42 & 0.078 &   2.32625 &  3879.1100 &  ASAS 182117-6638.8	      &   EA		   \\
UNSW-V-633  &   18 25 26.2 & $-$67 34 42.4 &	10.90 & 0.231 &   0.42713 &  3866.2150 &  ASAS 182528-6734.8	      &   EW		   \\
UNSW-V-639  &   18 30 46.4 & $-$67 08 15.2 &	12.43 & 0.293 &   1.85125 &  3886.2200 &  NSV 10858			      &   EA		   \\
UNSW-V-641  &   18 33 32.9 & $-$66 54 00.5 &	 9.35 & 0.011 &   1.92981 &  3867.2800 &  ASAS 183333-6654.0	      &   PUL		   \\
UNSW-V-643  &   18 34 13.5 & $-$66 07 10.0 &	11.98 & 0.340 &   -	  &  -     &  ASAS 183414-6607.2	      &   LPV		   \\
UNSW-V-645  &   18 35 38.8 & $-$66 55 52.6 &	 8.94 & 0.032 &   0.84381 &  3866.2300 &  ASAS 183540-6656.0	      &   EB		   \\
UNSW-V-646  &   18 36 25.9 & $-$67 56 03.1 &	 8.52 & 1.293 &   -	  &  -     &  DG PAV/ASAS 183627-6756.0       &   LPV		   \\
  \hline		     
\end{tabular}
}
\end{minipage}
\end{table*}

\begin{table*}
 \centering
 \begin{minipage}{140mm}
  \contcaption{}
 \label{tab:gcvs3}
{\scriptsize 
  \begin{tabular}{@{}lcccccclll@{}}
  \hline
  ID & RA         & Dec        & $I$ & $A$ & Period & Epoch & Alternate ID & Type\\
     & (J2000.0)  & (J2000.0)  & (mag) & (mag) & (d) & HJD-2450000.0 & & \\
  \hline
UNSW-V-656  &   21 04 47.3 & $-$66 46 16.6 &	10.02 & 0.032 &   0.42922 &  3937.2050 &  ASAS 210447-6646.3		      &   EW		 \\
UNSW-V-658  &   21 06 49.1 & $-$66 33 51.0 &	11.00 & 0.119 &   3.02652 &  3937.8800 &  ASAS 210649-6633.8	      &   EA		   \\
UNSW-V-665  &   21 11 35.9 & $-$66 12 49.8 &	12.46 & 0.152 &   0.37251 &  3937.1700 &  ASAS 211136-6612.8		      &   EW		 \\
UNSW-V-672  &   21 17 04.8 & $-$67 01 47.3 &	 8.97 & 0.129 &  45.25221 &	   -     &  ASAS 211705-6701.8	      &   LPV		   \\
UNSW-V-674  &   21 18 53.4 & $-$67 16 14.8 &	 8.48 & 0.051 &  41.88170 &	   -     &  ASAS 211855-6716.2	      &   LPV		   \\
UNSW-V-675  &   21 20 21.8 & $-$65 46 45.6 &	13.57 & 0.148 &   0.46078 &  3937.0900 &  NSV 13650			      &   RRL		   \\
UNSW-V-677  &   21 20 51.0 & $-$65 50 15.5 &	 8.84 & 0.033 &  32.99997 &  3046.0000 &  ASAS 212051-6550.3	      &   LPV		   \\
UNSW-V-690  &   21 02 58.2 & $-$68 45 12.8 &	10.66 & 0.166 &   0.51007 &  3937.3190 &  ASAS 210258-6845.2		      &   EW		 \\
UNSW-V-695  &   21 06 59.9 & $-$68 21 03.5 &	 8.88 & 0.052 &  -	  &   -        &  ASAS 210700-6821.0	      &   LPV		   \\
UNSW-V-714  &   23 48 48.7 & $-$69 46 53.4 &	11.32 & 0.101 &   0.39326 &  3991.2520 &  ASAS 234849-6946.9	      &   EW		   \\
UNSW-V-716  &   23 30 40.7 & $-$69 53 33.5 &	10.68 & 0.097 &   0.63150 &  3991.1660 &  ASAS 233041-6953.5	      &   EW		   \\
UNSW-V-719  &   23 33 56.1 & $-$69 11 14.1 &	11.20 & 0.101 &   0.95375 &  3990.9800 &  ASAS 233356-6911.2	      &   EW		   \\
UNSW-V-723  &   07 55 30.6 & $-$66 59 44.7 &	11.03 & 0.114 &   1.10830 &  4086.7700 &  ASAS 075530-6659.7	      &   EW		   \\
UNSW-V-724  &   07 55 14.7 & $-$68 08 11.9 &	10.85 & 0.196 &  97.58038 &	   -     &  ASAS 075514-6808.2	      &   LPV		   \\
UNSW-V-726  &   07 57 00.3 & $-$66 35 51.3 &	 9.32 & 0.052 &   1.29478 &  4087.1300 &  ASAS 075700-6635.8	      &   EB		   \\
UNSW-V-733  &   08 00 55.4 & $-$66 41 11.1 &	 9.96 & 0.023 &  46.64445 &	   -     &  ASAS 080055-6641.2	      &   LPV		   \\
UNSW-V-735  &   08 01 35.6 & $-$68 19 36.3 &	 8.14 & 0.097 &  66.31470 &	   -     &  ASAS 080135-6819.6	      &   LPV	 \\
UNSW-V-736  &   08 01 59.7 & $-$68 17 39.5 &	 9.14 & 0.198 &  94.34667 &	   -     &  ASAS 080200-6817.7	      &   LPV	      \\
UNSW-V-739  &   08 04 10.3 & $-$67 54 56.5 &	11.53 & 0.161 &   0.41287 &  4085.9850 &  ASAS 080410-6755.0	      &   EW		   \\
UNSW-V-741  &   08 04 53.8 & $-$66 48 49.3 &	11.19 & 0.102 &   0.48046 &  4086.0050 &  ASAS 080454-6648.8	      &   EW		   \\
UNSW-V-745  &   08 07 15.2 & $-$67 12 17.1 &	 9.21 & 0.027 &  28.64224 &  4110.0000 &  ASAS 080715-6712.3	      &   LPV		   \\
UNSW-V-755  &   08 12 59.1 & $-$67 14 44.3 &	11.61 & 0.124 &   0.34062 &  4085.9850 &  ASAS 081259-6714.7	      &   EW		   \\
UNSW-V-756  &   08 14 09.5 & $-$68 02 13.0 &	11.31 & 0.088 &   0.41789 &  4086.1350 &  ASAS 081409-6802.2	      &   EW		   \\
UNSW-V-758  &   08 16 09.3 & $-$66 44 46.3 &	11.94 & 0.139 &   0.38501 &  4086.1000 &  ASAS 081610-6644.8	      &   RRL	       \\
UNSW-V-761  &   07 52 35.7 & $-$67 15 11.1 &	 8.22 & 0.067 &   1.01055 &	   -     &  ASAS 075236-6715.2	      &   LPV		   \\
UNSW-V-768  &   14 36 15.8 & $-$69 51 10.3 &	12.34 & 0.165 &   -	  &  -     &  XZ CIR			      &   LPV		   \\
UNSW-V-770  &   14 39 01.4 & $-$69 22 43.2 &	11.15 & 0.168 &   -	  &  -     &  NSV 06732 		      &   LPV		   \\
UNSW-V-801  &   14 47 44.1 & $-$68 54 05.6 &	 9.57 & 0.072 &   7.24716 &  4175.8000 &  ASAS 144744-6854.1		      &   CEP		   \\
UNSW-V-806  &   14 49 45.1 & $-$69 35 31.7 &	 9.16 & 0.080 &   -	  &  -     &  ASAS 144945-6935.6	      &   LPV		   \\
UNSW-V-807  &   14 49 56.5 & $-$69 20 50.8 &	11.93 & 0.521 &   -	  &  -     &  BL CIR			      &   LPV		   \\
UNSW-V-824  &   14 56 41.9 & $-$68 34 47.9 &	10.92 & 0.129 &   1.05497 &  4170.2500 &  ASAS 145643-6834.8	      &   EA		   \\
UNSW-V-825  &   14 57 20.7 & $-$69 45 02.2 &	10.45 & 0.473 &   -	  &  -     &  ASAS 145720-6945.0	      &   LPV		   \\
UNSW-V-826  &   14 56 33.0 & $-$68 08 35.2 &	 9.47 & 0.212 &   2.13147 &  4173.1100 &  EM TRA/ASAS 145633-6808.6	      &   EA		   \\
UNSW-V-842  &   15 02 19.0 & $-$68 16 01.1 &	11.67 & 0.027 &   1.78869 &  4184.2700 &  NSV 06882			      &   EA		   \\
UNSW-V-844  &   15 03 08.6 & $-$69 50 58.3 &	11.29 & 0.192 &   0.36855 &  4170.0500 &  ASAS 150308-6950.9	      &   EW		   \\
UNSW-V-847  &   15 05 11.1 & $-$68 45 56.7 &	11.01 & 0.154 &   0.33718 &  4170.1500 &  ASAS 150511-6845.9	      &   EW		   \\
  \hline		     
\end{tabular}		     
}			     
\end{minipage}		     
\end{table*}

\begin{table*}
  \centering
  \begin{minipage}{140mm}
    \caption{UNSW variable stars coincident with ROSAT X-ray sources.}
    \label{tab:rosat}
    {\scriptsize 
      \begin{tabular}{@{}lcccccclll@{}}
      	\hline
      	ID & RA        & Dec       & $I$   & $A$   & Period & Epoch         & ROSAT ID & Type\\
      	   & (J2000.0) & (J2000.0) & (mag) & (mag) & (d)    & HJD-2450000.0 &          &     \\
      	\hline
UNSW-V-005 &   04 57 28.8 &$-$29 09 48.3 & 8.75 & 0.005 &    3.26837 &3289.1200 & 1RXS J045728.9-290953   &	  EB		     \\  
UNSW-V-040 &   09 05 22.3 &$-$15 03 42.9 & 9.60 & 0.007 &    0.62883 &3377.0800 & 1RXS J090522.2-150302   &	  PUL		     \\
UNSW-V-362 &   09 16 44.1 &$-$24 47 42.9 & 9.56 & 0.012 &    2.59901 &3744.1000 & 1RXS J091644.7-244735   &	  CEP  	      \\
UNSW-V-450 &   12 54 31.3 &$-$46 07 36.5 & 8.68 & 0.014 &    1.04272 &3805.1500 & 1RXS J125430.7-460735   &	  CEP  	      \\
UNSW-V-468 &   12 47 55.7 &$-$44 57 34.1 & 8.94 & 0.059 &    -       &- 	& 1RXS J124757.9-445735   &	  PUL		    \\
UNSW-V-470 &   12 48 07.6 &$-$44 39 17.5 & 8.59 & 0.053 &    1.04962 &3788.7300 & 1RXS J124807.6-443913   &	  CEP  \\
UNSW-V-493 &   13 13 07.4 &$-$45 37 30.3 & 9.56 & 0.013 &    6.85120 &3791.4000 & 1RXS J131306.7-453740   &	   CEP 	      \\
UNSW-V-494 &   13 13 20.7 &$-$45 38 13.4 & 8.21 & 0.115 &   48.60485 &   0.0000 & 1RXS J131306.7-453740   &	   LPV     \\
UNSW-V-506 &   13 16 38.9 &$-$45 46 56.0 & 8.88 & 0.003 &    0.08857 &3788.0200 & 1RXS J131651.3-454905   &	  PUL		    \\
UNSW-V-508 &   13 16 49.0 &$-$45 48 47.2 & 8.82 & 0.195 &    0.40969 &3788.1050 & 1RXS J131651.3-454905   &	  EW		     \\
UNSW-V-510 &   13 17 24.2 &$-$45 28 17.4 & 9.84 & 0.070 &   72.84351 &   0.0000 & 1RXS J131717.7-452541   &	  LPV		     \\
UNSW-V-514 &   13 17 46.5 &$-$44 56 39.3 &10.19 & 0.013 &    0.48342 &3788.0700 & 1RXS J131747.3-445707   &	  PUL		    \\
UNSW-V-521 &   13 22 04.2 &$-$45 03 10.8 & 9.03 & 0.022 &    1.44852 &3787.8500 & 1RXS J132204.7-450312   &	  CEP  	      \\
UNSW-V-525 &   14 44 14.2 &$-$39 10 15.9 &10.29 & 0.003 &    0.09340 &3847.0150 & 1RXS J144357.0-390847   &	   PUL  	\\
UNSW-V-538 &   14 40 47.7 &$-$38 47 05.7 & 9.21 & 0.007 &    0.19950 &3847.1700 & 1RXS J144037.4-384658   &	  PUL		    \\
UNSW-V-541 &   14 49 26.1 &$-$39 50 48.4 & 9.67 & 0.012 &    3.74574 &3847.3000 & 1RXS J144925.7-395042   &	  CEP  	      \\
UNSW-V-559 &   14 44 04.4 &$-$40 59 23.9 & 8.15 & 0.017 &    0.50383 &3846.9200 & 1RXS J144405.2-405940   &	  EW		     \\
UNSW-V-568 &   14 48 13.2 &$-$41 03 00.0 & 9.72 & 0.015 &    -       &- 	& 1RXS J144812.6-410310   &	  PUL		     \\
UNSW-V-577 &   14 52 42.1 &$-$41 41 55.3 & 9.60 & 0.032 &    0.87351 &3847.0400 & 1RXS J145240.7-414206   &	  RRL	\\
UNSW-V-582 &   14 42 16.0 &$-$41 00 19.0 & 9.43 & 0.020 &    2.57400 &3848.3000 & 1RXS J144214.5-410026   &	  CEP: 	     \\
UNSW-V-718 &   23 32 37.1 &$-$69 54 31.2 & 8.61 & 0.018 &    -       &-         & 1RXS J233239.5-695432   &       PUL                \\
UNSW-V-760 &   07 51 49.4 &$-$68 14 04.3 &10.70 & 0.026 &    0.19789 &4086.0500 & 2RXP J075145.0-681416   &       PUL               \\
      	\hline
      \end{tabular}
    }
  \end{minipage}
\end{table*}


\begin{thebibliography}{99}

\bibitem[\protect\citeauthoryear{Aigrain \& Irwin}{2004}]{Aigrain04} Aigrain S., Irwin A., 2004, MNRAS, 350, 331
\bibitem[\protect\citeauthoryear{Aigrain et al.}{2007}]{Aigrain07} Aigrain S., et al., 2007, MNRAS, 375, 29
\bibitem[\protect\citeauthoryear{Arbutina}{2007}]{Arbutina07} Arbutina B., 2007, MNRAS, 377, 1635
\bibitem[\protect\citeauthoryear{Bessell \& Brett}{1988}]{Bessell88} Bessell M., Brett J.~M., 1988, PASP, 100, 1134
\bibitem[\protect\citeauthoryear{Carpenter}{2001}]{Carpenter01} Carpenter J.~M., 2001, AJ, 121, 2851
\bibitem[\protect\citeauthoryear{Christiansen et al.}{2007}]{Christiansen07} Christiansen J.~L., Derekas A., Ashley M.~C.~B., Webb J.~K., Hidas M.~G., Hamacher D.~W., Kiss L.~L., 2007, MNRAS, 382, 239
\bibitem[\protect\citeauthoryear{Clarke}{2002}]{Clarke02} Clarke D., 2002, \aaa, 386, 763
\bibitem[\protect\citeauthoryear{Clement et al.}{2006}]{Clement06} Clement C.~M., Nguyen D.~C., Rucinski S.~M., Yee H.~K., Mallen-Ornelas G., Gladders M.~D., Seager S., 2006, \aas, 208, 4408
\bibitem[\protect\citeauthoryear{Collier Cameron et al.}{2007}]{Cameron07} Collier Cameron A., et al., 2007, MNRAS, 375, 951
\bibitem[\protect\citeauthoryear{Covino}{2004}]{Covino04} Covino E., Frasca A., Alcal\'a J.M., Paladino R., Sterzik M.F., 2004, \aaa, 427, 637
\bibitem[\protect\citeauthoryear{Derekas et al.}{2007}]{Derekas07} Derekas A., Kiss L.~L., Bedding T.~R., 2007, ApJ, 663, 249
\bibitem[\protect\citeauthoryear{Hartman et al.}{2004}]{Hartman04} Hartman J.~D., Bakos G., Stanek K.~Z., Noyes R.~W., 2004, AJ, 128, 1761
\bibitem[\protect\citeauthoryear{Hebb et al.}{2006}]{Hebb06} Hebb L., Wyse R.~F.~G., Gilmore G., Holtzman J., 2006, AJ, 131, 555
\bibitem[\protect\citeauthoryear{Hidas et al.}{2005}]{Hidas05} Hidas M.~G. et al., 2005, MNRAS, 360, 703
\bibitem[\protect\citeauthoryear{Irwin \& Lewis}{2001}]{Irwin01} Irwin M., Lewis J., 2001, New Astronom. Rev, 45, 105
\bibitem[\protect\citeauthoryear{Irwin et al.}{2007}]{Irwin07} Irwin J., et al., 2007, MNRAS, 380, 541
\bibitem[\protect\citeauthoryear{Jurcsik et al.}{2006}]{Jurcsik06} Jurcsik J., et al., 2006, \aj, 132, 61
\bibitem[\protect\citeauthoryear{Kolenberg et al.}{2006}]{Kolenberg06} Kolenberg K., et al., 2006, \aaa, 459, 577
\bibitem[\protect\citeauthoryear{Kov\'acs}{2001}]{Kovacs01} Kov\'acs G., 2001, in Stellar pulsation - nonlinear studies, eds. Takeuti M., \& Sasselov D.D., ApSS Library (Dordrecht: Kluwer Academic Publishers), 257, 61
\bibitem[\protect\citeauthoryear{Kov{\' a}cs, Bakos \& Noyes}{2005}]{Kovacs05} Kov\'acs G., Bakos G., Noyes R.~W., 2005, MNRAS, 356, 557
\bibitem[\protect\citeauthoryear{Lafler \& Kinman}{1965}]{Lafler65} Lafler J., Kinman T.~D., 1965, ApJS, 11, 216
\bibitem[\protect\citeauthoryear{Li \& Zhang}{2006}]{Li06} Li L., Zhang F., 2006, MNRAS, 369, 2001
\bibitem[\protect\citeauthoryear{Makarov}{2003}]{Makarov03} Makarov V.~V., 2003, AJ, 126, 1996
\bibitem[\protect\citeauthoryear{Mapelli et al.}{2004}]{Mapelli04} Mapelli M., et al., 2004, ApJ, 605, L29
\bibitem[\protect\citeauthoryear{McDowell}{2007}]{McDowell07} McDowell J., 2007 IVOA Recommendation 29 October 2007, \url{http://www.ivoa.net/Documents/latest/SpectrumDM.html}
\bibitem[\protect\citeauthoryear{Norton et al.}{2007}]{Norton07} Norton A.~J. et al., 2007, {\bf \aaa, 467, 785}
\bibitem[\protect\citeauthoryear{Ochsenbein et al.}{2000}]{Ochsenbein00} Ochsenbein F., Bauer P., Marcourt J., 2001, \aas, 143, 23
\bibitem[\protect\citeauthoryear{Ochsenbein et al.}{2004}]{Ochsenbein04} Ochsenbein F., 2004, IVOA Recommendation 11 August 2004, \url{http://www.ivoa.net/Documents/latest/VOT.html}
\bibitem[\protect\citeauthoryear{Ol\`ah et al.}{2000}]{Olah00} Ol\'ah K., Koll\'ath Z., Strassmeier K.G., 2000, \aaa, 356, 643
\bibitem[\protect\citeauthoryear{Pepper \& Burke}{2006}]{Pepper06} Pepper J., Burke C., 2006, AJ, 132, 1177
\bibitem[\protect\citeauthoryear{Pojmanski}{2002}]{Pojmanski02} Pojmanski G., 2002, Acta Astron., 52, 397
\bibitem[\protect\citeauthoryear{Pribulla et al.}{1999}]{Pribulla99} Pribulla T., Chochol D., Rovithis-Livanou H., Rovithis P., 1999, \aaa, 345, 137
\bibitem[\protect\citeauthoryear{Rasio}{1995}]{Rasio95} Rasio F.~A., 1995, ApJ, 444, L41
\bibitem[\protect\citeauthoryear{ROSAT}{2000}]{ROSAT00} ROSAT Consortium, 2000, ROSAT News, 72
\bibitem[\protect\citeauthoryear{Rucinski}{2006}]{Rucinski06} Rucinski S., 2006, MNRAS, 368, 1319
\bibitem[\protect\citeauthoryear{Rucinski}{2007}]{Rucinski07} Rucinski S., 2007, MNRAS, 382, 393
\bibitem[\protect\citeauthoryear{Samus et al.}{2007}]{Samus07} Samus N.N., et al., 2007, The combined table of General Catalogue of Variable Stars Vols. I-III, 4th. ed. and Namelists of Variable Stars Nos.~67-78 with improved coordinates, Sternberg Astronomical Insitute, Moscow
\bibitem[\protect\citeauthoryear{Skrutskie et al.}{2006}]{Skrutskie06} Skrutskie M.~F., et al., 2006, AJ, 131, 1163 
\bibitem[\protect\citeauthoryear{Soszy{\' n}ski}{2006}]{Soszynski06} Soszy{\' n}ski I., 2006, MmSAI, 77, 265
\bibitem[\protect\citeauthoryear{Stassun et al.}{2004}]{Stassun04} Stassun K.~G., Mathieu R.~D., Vaz L.~P.~R., Stroud N., Vrba F.~J., 2004, ApJS, 151, 357
\bibitem[\protect\citeauthoryear{Stassun et al.}{2007}]{Stassun07} Stassun K.~G., Mathieu R.~D., Valenti J.~A., 2007, ApJ, 664, 1154
\bibitem[\protect\citeauthoryear{Stepien}{2007}]{Stepien07} Stepien K., 2006, Acta Astron., 56, 347
\bibitem[\protect\citeauthoryear{Stetson}{1996}]{Stetson96} Stetson P.~B., 1996, PASP, 108, 851
\bibitem[\protect\citeauthoryear{Stothers}{2006}]{Stothers06} Stothers R.~B., 2006, ApJ, 652, 643
\bibitem[\protect\citeauthoryear{Szalai et al.}{2007}]{Szalai07} Szalai T., Kiss L.L., M\'esz\'aros Sz., Vink\'o J., Csizmadia Sz., 2007, \aaa, 465,943
\bibitem[\protect\citeauthoryear{Szczygiel \& Fabrycky}{2007}]{Szczygiel07} Szczygiel D.~M., Fabrycky D.~C., 2007, MNRAS, 377, 1263
\bibitem[\protect\citeauthoryear{Tamuz, Mazeh \& Zucker}{2005}]{Tamuz05} Tamuz O., Mazeh T., Zucker S., 2005, MNRAS, 356, 1466
\bibitem[\protect\citeauthoryear{Taylor}{2005}]{Taylor05} Taylor M., 2005, ASPC, 347, 29
\bibitem[\protect\citeauthoryear{Tody \& Plante}{2004}]{Tody04} Tody D., Plante R., 2004, IVOA Working Draft 24 May 2004, \url{http://www.ivoa.net/Documents/latest/SIA.html}
\bibitem[\protect\citeauthoryear{Tody et al.}{2007}]{Tody07} Tody D., et al., 2007, IVOA Proposed Recommendation 17 September 2007, \url{http://www.ivoa.net/Documents/latest/SSA.html}
\bibitem[\protect\citeauthoryear{Voges et al.}{1999}]{Voges99} Voges W., et al., 1999, \aaa, 349, 389
\bibitem[\protect\citeauthoryear{Voges et al.}{2000}]{Voges00} Voges W., et al., 2000, IAUC 7432
\bibitem[\protect\citeauthoryear{Weldrake et al.}{2004}]{Weldrake04} Weldrake D.~T.~F., Sackett P.~D., Bridges T.~J., Freeman K.~C., 2004, AJ, 128, 736
\bibitem[\protect\citeauthoryear{Weldrake et al.}{2007}]{Weldrake07} Weldrake D.~T.~F., Sackett P.~D., Bridges T.~J., 2007, AJ, 133, 1447
\bibitem[\protect\citeauthoryear{Young et al.}{2007}]{Young07} Young T.~B., Hidas M.~G., Webb J.~K., Ashley M.~C.~B., Christiansen J.~L., Derekas A., Nutto C., 2006, MNRAS, 370, 1529


\end{thebibliography}
\end{document}